\documentclass[12pt]{article}
\usepackage{jheppub}
\usepackage{graphicx}
\usepackage{bm}  
\usepackage{slashed}
\usepackage{subfig}
\newcommand\beq{\begin{equation}}
\newcommand\eeq{\end{equation}}

\newcommand\hc{\text{h.c.}}

\abstract{We analyze the Type-II two Higgs doublets model in light of
  the newly discovered Higgs-like particle with mass 125 GeV.  The
  observed 125 GeV particle is identified with the light CP-even Higgs
  boson in the two Higgs doublets model.  We study the parameter space
  of the model consistent with the Higgs data, branching ratio of
  $\bar{B}\to X_s\gamma$ as well as precision electroweak
  measurements.  We also incorporate theoretical constraints---
  perturbativity of the couplings and vacuum stability, in our study.
  We find that only a small parameter space of the model remains
  viable. The phenomenology of the heavy Higgs bosons in the surviving
  parameter space is studied.  }

\begin{document}

\title{Carving Out Parameter Space in Type-II Two Higgs Doublets Model}

\author[a]{Benjam\'\i{}n Grinstein,}
\author[b,c]{Patipan Uttayarat}

\affiliation[a]{Department of Physics, University of California at San Diego, La Jolla, CA 92093}
\affiliation[b]{Department of Physics, University of Cincinnati, Cincinnati, OH 45220}
\affiliation[c]{Department of Physics, Srinakharinwirot University, Wattana, Bangkok 10110 Thailand}

\emailAdd{bgrinstein@ucsd.edu}
\emailAdd{uttayapn@ucmail.uc.edu}

\preprint{UCSD/PTH 13-02}

\maketitle

\section{Introduction}
Recently the ATLAS and CMS collaborations announced the discovery of a
Higgs-like particle with a mass $M_h\simeq125$
GeV~\cite{:2012gk,:2012gu}.  Evidence of this new particle has also
been reported by the CDF and D$\O$
collaborations~\cite{Aaltonen:2012qt}.  However, it is far from certain that this newly
discovered particle is the standard model (SM) Higgs boson responsible
for electroweak symmetry breaking.  The couplings of this
Higgs-like particle deviate (although not statistically significantly)
from SM expectations~\cite{Carmi:2012in,Espinosa:2012im,Corbett:2012ja,Azatov:2012qz}.  Even with
the updated
measurements~\cite{ATLAS-CONF-2012-091,ATLAS-CONF-2012-092,ATLAS-CONF-2012-158,CMS-PAS-HIG-12-045,TEVNPH:2012ab,ATLAS-CONF-2013-013,CMS-PAS-HIG-13-002,CMS-PAS-HIG-13-003},
the deviations still remain.  Thus it is possible that this Higgs-like
particle is a hint of new physics beyond the standard model.

A particularly well motivated class of new physics is the two Higgs
doublet model in which electroweak symmetry is broken by two
elementary scalar fields.  Famously, the
minimal supersymmetric standard model (MSSM) contains two higgs
doublets to account for masses of all quarks and leptons, and the
parameters are constrained by supersymmetry.  However, in this
work we consider  a more generic two Higgs doublet model.
For a recent review of a general two Higgs doublet model see
Ref.~\cite{Branco:2011iw} and references therein.  In particular we
will focus on the CP conserving type II two Higgs doublet model
(2HDM-II) in which one scalar field couples only to the up-type quarks
and the other couples to the down-type quarks and leptons.

There exists a large literature on the properties of the neutral
scalar boson couplings of the two Higgs doublets model in light of the Higgs data; 
see for example Ref.~\cite{Cheon:2012rh,Bai:2012ex,Altmannshofer:2012ar,Drozd:2012vf,Chang:2012zf,Chen:2013kt,Goudelis:2013uca}.
In this work, in addition to performing a global fit to the currently
available Higgs data, we also consider the viable parameter space of
the model and  study the phenomenology of the other Higgs bosons.\footnote{Ref.~\cite{Cheon:2012rh} also studied the viable parameter space of the 2HDM-II but didn't discuss the phenomenology of the other Higgs bosons.}  
Specifically, we assume that there is no other states
except those of the 2HDM-II up to some cutoff scale, $\Lambda$.  Thus
if there is no viable parameter space for a specific value of the
cutoff, we can conclude that, if there is new physics beyond the standard
model, the 2HDM-II cannot be the only new physics below that cutoff scale.

From theoretical view point, the model has to allow for an electroweak
symmetry breaking vacuum.  We also impose a constraint on
perturbativity of the coupling constants of the model.  What we mean
by perturbativity will be made clear in
section~\ref{subsec:theorybounds}.  We insist that perturbativity must
be satisfied at all energy scales up to the cutoff
scale~\cite{Ferreira:2009jb}.  This is different from
Ref~\cite{Drozd:2012vf} which seems to impose perturbativity only at
the electroweak scale.

Existing experimental data also constrain possible new physics.  The
absence of large flavor changing neutral interactions places a strong
bound on the mass of the charged Higgs boson.  Similarly, the success
of the standard model in describing precision electroweak measurements
constrains possible new physics states.  In this work, we will
utilize both experimental and theoretical constraints in determining a
viable parameter space for the 2HDM-II.

The paper is organized as follows.  In section~\ref{sec:model} we
briefly describe the 2HDM-II and we set our conventions and
notations.  Then we perform a global fit to the Higgs data in
section~\ref{sec:fit}.  In section~\ref{sec:bounds} we subject the
2HDM-II to both theoretical and experimental constraints to determine
the viable parameter space.  We briefly discuss the phenomenology of
the heavy CP-even neutral Higgs boson, $H$, in the surviving parameter
space in section~\ref{sec:pheno}.  We then conclude in
section~\ref{sec:conclusion}.


\section{The Model}
\label{sec:model}
Here we give a brief overview of the 2HDM-II and we set our
notation.  We take the two scalar doublets $\Phi_1$ and $\Phi_2$ to
have hypercharge $1/2$. They can be expanded as
\begin{equation}
	\Phi_j = \begin{pmatrix}\phi_j^+\\(v_j+\rho_j+i\eta_j)/\sqrt{2}\end{pmatrix}.
\end{equation}

The Yukawa coupling to fermions are given by
\begin{equation}
	\mathcal{L}_{yuk} = -y_u\bar{q}_L(i\sigma^2\Phi_2^\ast)u_R - y_d\bar{q}_L\Phi_1d_R - y_e\bar{e}_L\Phi_1e_R +\hc
\end{equation}
Expanding out the CP-even neutral scalar sector we obtain
\begin{equation}
	\mathcal{L}_{yuk} = -\frac{M_u}{v\sin\beta}\bar u u\rho_2 - \frac{M_d}{v\cos\beta}\bar d d\rho_1 - \frac{M_e}{v\cos\beta}\bar e e\rho_1,
\end{equation}
where $\tan\beta = v_2/v_1$ and $v^2=v_1^2+v_2^2$.  The two CP-even
neutral scalars mix with each other.  The mass eigenstates are given
by
\begin{equation}
	\begin{pmatrix}\rho_1\\ \rho_2\end{pmatrix} = \begin{pmatrix}\cos\alpha & \sin\alpha\\ -\sin\alpha &\cos\alpha\end{pmatrix}\begin{pmatrix}H \\ h\end{pmatrix},
\end{equation}
where $h$ is the lighter eigenstate to be identified with the observed
125~GeV Higgs-like particle. In terms of these mass eigenstates, we
find
\begin{equation}
\label{eq:yukawacouplings}
\begin{split}
	\mathcal{L}_{yuk} &= -\frac{M_u}{v}\left(\frac{\cos\alpha}{\sin\beta}\right) \bar u u h - \frac{M_d}{v}\left(\frac{\sin\alpha}{\cos\beta}\right) \bar d d h - \frac{M_e}{v}\left(\frac{\sin\alpha}{\cos\beta}\right) \bar e e h \\
	&\quad + \frac{M_u}{v}\left(\frac{\sin\alpha}{\sin\beta}\right) \bar u u H - \frac{M_d}{v}\left(\frac{\cos\alpha}{\cos\beta}\right) \bar d d H - \frac{M_e}{v}\left(\frac{\cos\alpha}{\cos\beta}\right) \bar e e H
\end{split}
\end{equation}

The couplings of these two eigenstate $h$ and $H$ to gauge bosons can
be obtained from the kinetic terms of $\Phi_1$ and $\Phi_2$.  They are
\begin{equation}
\begin{aligned}
	\frac{2M_W^2}{v}&W_\mu^+W^{-\mu}h\sin(\alpha+\beta)+\frac{M_Z^2}{v}Z_\mu Z^\mu h\sin(\alpha+\beta)\\
	&+\frac{2M_W^2}{v}W_\mu^+W^{-\mu}H\cos(\alpha+\beta)+\frac{M_Z^2}{v}Z_\mu Z^\mu H\cos(\alpha+\beta).
\end{aligned}
\end{equation}

\subsection{Scalar Sector}
\label{sec:scalarsector}
The scalar sector of the 2HDM is the most model dependent part.  Here
we will focus on the simplest scalar potential consistent with CP
symmetry
\begin{equation}
\label{eqn:scalarpotential}
\begin{split}
	V(\Phi_1,\Phi_2) &= m_{11}^2|\Phi_1|^2 + m_{22}^2|\Phi_2|^2 +\frac{\lambda_1}{2}\left(|\Phi_1|^2\right)^2 + \frac{\lambda_2}{2}\left(|\Phi_2|^2\right)^2 + \lambda_3|\Phi_1|^2|\Phi_2|^2\\
	&\qquad + \lambda_4|\Phi_1^\dagger\Phi_2|^2+\frac{\lambda_5}{2}\left[\left(\Phi_1^\dagger\Phi_2\right)^2+\left(\Phi_2^\dagger\Phi_1\right)^2\right]
\end{split}
\end{equation}

It is more convenient to characterize the scalar sector by their
physical masses and the mixing angles--- $M_h$,
$M_H$, $M_A$, $M_{H^\pm}$, $\alpha$, $\tan\beta$:
\begin{equation}
\label{eq:scalarcouplings}
\begin{split}
	\sin^2\alpha\, M_h^2 + \cos^2\alpha\, M_H^2 &=\frac{v^2}{1 + \tan^2\beta} \lambda_1,\\ 
	\cos^2\alpha\, M_h^2 + \sin^2\alpha\, M_H^2 &= \frac{\tan^2\beta}{1 + \tan^2\beta}v^2 \lambda_2,\\
	(M_h^2-M_H^2)\cos\alpha\sin\alpha &= (\lambda_3+\lambda_4+\lambda_5)\frac{\tan\beta}{1+\tan^2\beta}v^2, \\
	M_A^2 &= -\lambda_5 v^2,\\
	M_{H^\pm}^2 &= -\frac{1}{2}(\lambda_4+\lambda_5)v^2 = M_A^2+\frac12(\lambda_5-\lambda_4)v^2.
\end{split}
\end{equation}
The set of parameters we use in our parameter-space scan consists of
$\alpha$, $\tan\beta$, $M_h$, $M_H$, $M_A$ and $M_{H^\pm}$. 
We identify the light CP-even scalar, h, with the observed 125GeV resonance. 
We do not discuss the alternative hypothesis, that the heavy CP- even scalar is identified with the 125 GeV resonance. 
The fit to higgs data cannot distinguish between these hypothesis because the couplings of $H$ are the same as those for $h$ after $\alpha\to\alpha+\pi/2$. 
Hence, for the remaining part of this work we set $M_h = 125$ GeV.

The scalar potential in equation~\eqref{eqn:scalarpotential} posses a discrete $Z_2$ symmetry forbidding terms with odd power of $\Phi_1$ or $\Phi_2$.  
This is the defining symmetry of the type II model, designed to avoid flavor changing neutral interactions at tree level~\cite{Glashow:1976nt}. It is a discrete symmetry, rather than continuous Peccei-Quinn U(1)-symmetry~\cite{Peccei:1977ur}, to avoid a light axion~\cite{Weinberg:1977ma,Wilczek:1977pj}.
It is conceivable that the symmetry is broken softly by adding to the potential the term $m_{12}^2\Phi_1^\dagger\Phi_2 +\hc.$
Such a term would add $m_{12}^2\{\tan\beta$, $\cot\beta$, -1, $1/(\sin\beta\cos\beta)$, $1/(\sin\beta\cos\beta)\}$, respectively, to the relations in equation~\eqref{eq:scalarcouplings}.
We will not pursue this possibility in this paper but we will discuss briefly the effect of this term on the viable parameter space in section~\ref{sub:paramspace}.

\section{Fit to the Higgs Data}
\label{sec:fit}
The mass parameters $M_H$, $M_A$ and $M_{H^\pm}$ affect Higgs data
observables only through suppressed radiative corrections. We determine
the VEV ratio $\tan\beta$ and the neutral scalar mixing angle $\alpha$
from all the reported Higgs data.

Experimental data are reported in terms of a signal strength, $\mu$, defined as
\begin{equation}
	\mu \equiv \frac{\sigma}{\sigma^{SM}}\frac{Br}{Br^{SM}}.
\end{equation}
When the signal strength is not directly reported by the experimental
collaboration, we extract it from the reported 95\% exclusion limit
following the procedure given in Ref.~\cite{Giardino:2012ww}. Here we
briefly review the procedure and refer the reader to the reference for
details.\footnote{A more refined procedure for extracting the signal strength was formulated in Ref.~\cite{Azatov:2012bz}. The two methods  give comparable results.}  The experiments report the upperbound on the rate at 95\%
C.L., $R_{obs}$, and the expected upperbound at 95\% C.L. in the
absence of the Higgs boson, $R_{exp}$.  The signal strength and its
uncertainty, $\sigma$, are given by
\begin{equation}
	\mu = R_{obs} - R_{exp},\qquad \sigma = \frac{R_{obs}}{1.96}.
\end{equation}
We collect the signal strengths for each search channel in Table \ref{tab:mu}.
\begin{table}[tc]
\begin{center}
\begin{tabular}{| l | c | c |}
\hline
Channel & Signal Strength ($\mu$)& Reference\\\hline
ATLAS $\gamma\gamma$, 7 TeV  &$1.6^{+0.9}_{-0.8}$ & \cite{ATLAS:2012ad}\\\hline
CMS $\gamma\gamma$, dijet-tagged, 7 TeV  &$4.21\pm2.04$ & \cite{CMS-PAS-HIG-12-015} \\\hline
CMS $\gamma\gamma$, untagged 0, 7 TeV  &$3.15\pm1.82$ & \cite{CMS-PAS-HIG-12-015}\\\hline
CMS $\gamma\gamma$, untagged 1, 7 TeV  &$0.66\pm0.95$ & \cite{CMS-PAS-HIG-12-015}\\\hline
CMS $\gamma\gamma$, untagged 2, 7 TeV  &$0.73\pm1.15$ & \cite{CMS-PAS-HIG-12-015}\\\hline
CMS $\gamma\gamma$, untagged 3, 7 TeV  &$1.53\pm1.61$ & \cite{CMS-PAS-HIG-12-015}\\\hline
CMS $\gamma\gamma$, dijet-tight, 8 TeV & $1.32\pm1.57$& \cite{CMS-PAS-HIG-12-015}\\\hline
CMS $\gamma\gamma$, dijet-loose, 8 TeV &$-0.61\pm2.03$ & \cite{CMS-PAS-HIG-12-015}\\\hline
CMS $\gamma\gamma$, untagged 0, 8 TeV  &$1.46\pm1.24$ & \cite{CMS-PAS-HIG-12-015}\\\hline
CMS $\gamma\gamma$, untagged 1, 8 TeV  &$1.51\pm1.03$ & \cite{CMS-PAS-HIG-12-015}\\\hline
CMS $\gamma\gamma$, untagged 2, 8 TeV  &$0.95\pm1.15$ & \cite{CMS-PAS-HIG-12-015}\\\hline
CMS $\gamma\gamma$, untagged 3, 8 TeV  &$3.78\pm1.77$ & \cite{CMS-PAS-HIG-12-015}\\\hline
ATLAS $\gamma\gamma$, 8 TeV  & $1.6\pm0.32$& \cite{ATLAS-CONF-2013-013}\\\hline
ATLAS ZZ, 7 TeV & $1.4^{+1.3}_{-0.8}$ &\cite{ATLAS:2012ac}\\\hline
CMS ZZ, 7 TeV & $0.6^{+0.9}_{-0.6}$& \cite{CMS-PAS-HIG-12-008}\\\hline
ATLAS ZZ, combine 7 \& 8 TeV & $1.5\pm0.6$ &\cite{ATLAS-CONF-2013-013}\\\hline
CMS ZZ, combine 7 \& 8 TeV & $0.91^{+0.30}_{-0.24}$ & \cite{CMS-PAS-HIG-13-002}\\\hline
ATLAS WW, 7 TeV & $0.1^{+0.7}_{-0.6}$ & \cite{ATLAS-CONF-2012-012}\\\hline
CMS WW, 7 TeV & $0.4\pm0.6$ & \cite{CMS-PAS-HIG-12-008}\\\hline
ATLAS WW, 8 TeV & $1.45\pm0.56$&\cite{ATLAS-CONF-2012-158}\\\hline
CMS WW, combine 7 \& 8 TeV & $0.76\pm0.21$ & \cite{CMS-PAS-HIG-13-003} \\\hline
CMS $b\bar{b}$, 7 TeV &$1.2^{+2.1}_{-1.7}$ & \cite{CMS-PAS-HIG-12-008}\\\hline
CMS $b\bar{b}$, 8 TeV &$1.07\pm0.62$ & \cite{CMS-PAS-HIG-12-045}\\\hline
Tevatron $b\bar{b}$ &$2.0\pm0.7$ & \cite{TEVNPH:2012ab}\\\hline
ATLAS $\tau\bar{\tau}$, 8 TeV &$0.7\pm0.7$ & \cite{ATLAS-CONF-2012-160}\\\hline
CMS $\tau\bar{\tau}$, 8 TeV &$0.88\pm0.50$ & \cite{CMS-PAS-HIG-12-045}\\\hline
\end{tabular}
\caption{The signal strengths and the corresponding error for the Higgs data used in the fit.}
\label{tab:mu}
\end{center}
\end{table}
\subsection{Production Cross-sections and Branching Ratios}
Here we work out the Higgs production cross-section and branching
ratios for a non-stadard model Higgs coupling.  At tree-level, the
non-standard Higgs couplings to SM fields can be characterized by
rescaling coefficients, $c_i$'s, relative to the standard model higgs
couplings as follows
\begin{equation}
\begin{split}
	\mathcal{L}_h &= c_V\frac{h}{v}\left(2m_W^2 W^+_\mu W^-_\mu+m_Z^2 Z_\mu Z_\mu\right) - c_t\frac{h}{v}m_t\bar{t}t - c_b\frac{h}{v}m_b\bar{b}b - c_c\frac{h}{v}m_c\bar{c}c - c_\tau\frac{h}{v}m_\tau\bar{\tau}\tau 
\end{split}
\end{equation}
In this work we assume that other scalars are sufficiently heavy that
their effects on the 125 GeV Higgs boson decay channels are
negligible.

The main production channels considered here are gluon fusion (ggF),
vector boson fusion (VBF), vector boson associated production (Vh) and
$t\bar{t}h$ production.  These production cross-sections are given in
terms of the SM ones by
\begin{equation}
	\frac{\sigma_{ggF}}{\sigma_{ggF}^{SM}} = \left|\frac{1.03c_t-0.05c_b}{1.03-0.05}\right|^2,\qquad
	\frac{\sigma_{VBF}}{\sigma_{VBF}^{SM}} = \frac{\sigma_{Vh}}{\sigma_{Vh}^{SM}} = \left|c_V\right|^2,\qquad
	\frac{\sigma_{t\bar{t}h}}{\sigma_{t\bar{t}h}^{SM}} = \left|c_t\right|^2
\end{equation}
The rescaling factors for the partial decays widths are
\begin{equation}
\begin{split}
	\Gamma_{bb} = |c_b|^2\Gamma_{bb}^{SM},\quad \Gamma_{cc} = |c_c|^2\Gamma_{cc}^{SM},\quad \Gamma_{\tau\tau} = |c_\tau|^2\Gamma_{\tau\tau}^{SM},\quad \Gamma_{VV} = |c_V|^2\Gamma_{VV}^{SM},\quad\\
	\frac{\Gamma_{gg}}{\Gamma_{gg}^{SM}} = \left|\frac{1.03c_t-0.05c_b}{1.03-0.05}\right|^2,\quad \frac{\Gamma_{\gamma\gamma}}{\Gamma_{\gamma\gamma}^{SM}} = \left|\frac{\frac29\,1.03c_t-1.04c_V}{\frac29\,1.03-1.04}\right|^2, \hspace{1.5cm}
\end{split}	
\end{equation}
where we have assumed that the loop-induced, charged Higgs contribution to
$\Gamma_{\gamma\gamma}$ is negligible.  This assumption can be
justified once we include experimental constraints from
section~\ref{sec:expconstraints}.  The rescaling factors for the
2HDM-II are given by
\begin{equation}
\begin{split}
	c_t = c_c = \frac{\cos\alpha}{\sin\beta},\quad
	c_b = c_\tau = \frac{\sin\alpha}{\cos\beta},\quad
	c_V = \sin(\alpha+\beta).
\end{split}
\end{equation}

For completeness we include the rescaling of the partial decay width
for $h\to\gamma Z$
\begin{equation}
	\frac{\Gamma_{\gamma Z}}{\Gamma_{\gamma Z}^{SM}} = \left|\frac{-(0.38+0.37i)0.86c_t-(3.32+3.98i)c_V}{-(0.38+0.37i)0.86-(3.32+3.98i)}\right|^2.
\end{equation}
The general expressions for the loop-induced decay of a neutral scalar
boson are collected in appendix~\ref{app:decays}.

\subsection{Global Fit}
\label{subsec:fit}

\begin{figure}
\begin{center}
	\includegraphics[width=0.7\textwidth]{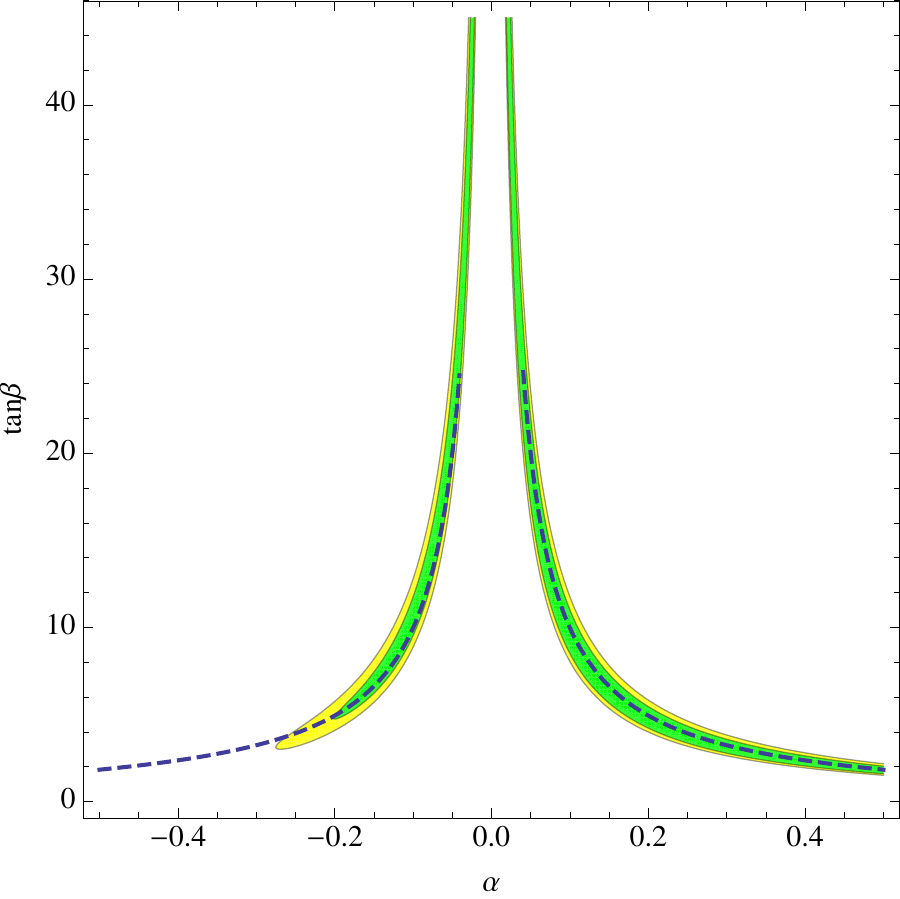}
	\caption{The region of parameter space within 1- and
          2-$\sigma$ of the best fit values.  The dashed line is the
          decoupling limit, $\alpha+\beta = \pm\pi/2$, where the
          couplings are SM-like (up to a possible sign flip for the
          down Yukawa couplings).}
	\label{fig:avstanb}
\end{center}
\end{figure}

We determine the best fit values of $\tan\beta$ and $\alpha$
using all the reported Higgs data from the Tevatron, LHC 7 TeV and LHC
8 TeV runs.  The theoretical SM Higgs boson predictions for the
cross-sections and branching ratios are taken from the LHC Higgs
Cross-section Working
Group~\cite{LHCHiggsCrossSectionWorkingGroup:2011ti}. We obtain  best
fit values  $\tan\beta \simeq 0.01$ and $\alpha \simeq
1.56$.  At this small value of $\tan\beta$ the top Yukawa coupling
is non-perturbative at the weak scale.  Thus the best fit value
doesn't seem to correspond to viable model parameters.

Assuming a Gaussian distribution around the best fit values, we
determine the region of parameter-space consistent within 1- and
2-$\sigma$ with the best fit values.  The result is shown in
figure~\ref{fig:avstanb}. The dashed line indicates the decoupling
limit, $\alpha+\beta = \pi/2$, where the couplings are SM-like (up to
a possible sign flip for the down Yukawa couplings). The figure is
truncated at $\alpha<0.5$, since for larger values $\tan\beta$
decreases, forcing the top Yukawa coupling into a non-perturbative
regime.



\section{Bounds on Parameter Space}
\label{sec:bounds}
\subsection{Theoretical Constraints}
\label{subsec:theorybounds}
We impose constraints amounting to the potential being bounded from
below. To this end the couplings must satisfy~\cite{Ferreira:2009jb}
\begin{equation}
\begin{split}
	\lambda_1, \lambda_2 &> 0\,,\\
	\lambda_3 &> -\sqrt{\lambda_1\lambda_2}\,,\\
	\lambda_3+\lambda_4-|\lambda_5|& > -\sqrt{\lambda_1\lambda_2}\,,
\end{split}
\end{equation}
at all scales up to the cutoff scale $\Lambda$.

We also impose the following perturbativity constraint on the couplings
\begin{equation}
\label{eq:perturbativity}
	\frac{y_i^2}{4\pi} \le 1\,,\qquad \frac{\lambda_i}{4\pi} \le 1.
\end{equation} 
We insist on these constraints up to the cutoff scale for all the
Yukawa and scalar couplings.  We list the beta-functions used in
evolving the coupling constants in appendix~\ref{app:beta}.

\subsection{Experimental Bounds}
\label{sec:expconstraints}
A wealth of experimental data, particularly from precision measurements,
places strong constraints on the spectrum of the 2HDM-II.  A newly
published result on a direct search for the charged Higgs at LEP
yields the 95\% CL lower bound $M_{H^\pm}\ge80$
GeV~\cite{Abbiendi:2013hk}.  At present there is no lower bound on
the charged Higgs mass from the Tevatron or LHC data.  A much tighter
constraint on the charged Higgs mass can be deduced from rare decay
processes.  By analyzing the branching ratio $Br(\bar{B}\to
X_s\gamma)$, Ref.~\cite{Hermann:2012fc} obtained the bound
$M_{H^\pm}\ge 380$ GeV at 95\% confidence level.  
A direct search at
LEP places a 95\% limit $M_A\gtrsim93$ GeV for the MSSM CP-odd Higgs, $A$~\cite{Schael:2006cr}. 
However this limit doesn't apply to the 2HDM case studied here.
Nevertheless, we employed this bound in in the rest of the paper.
The reader should keep in mind that $M_A\lesssim93$ GeV is not experimentally excluded.

Electroweak precision measurements also place strong constraints on
the spectrum of a 2HDM.  We concentrate on the oblique $S$
and $T$ parameters. In the Standard Model, the best fit values for the $S$ and $T$ parameters for $M_h\sim125$ GeV and $M_t=173$
GeV, as well as their correlation matrix ($M_{corr}$) are~\cite{Baak:2012kk}.   
\begin{equation}
\begin{aligned}
	S &= 0.03 \pm 0.10,\\
	T &= 0.05 \pm 0.12,  
\end{aligned}
\qquad
\qquad
	M_{corr} = 
\begin{pmatrix}
	1\quad & 0.891\\
	0.891 &1
\end{pmatrix}.
\end{equation}
Additional contribution to the $S$ and $T$ parameters from the heavy Higgs bosons are given in Ref.~\cite{Branco:2011iw}.

\subsection{Viable Parameter Space}
\label{sub:paramspace}

\begin{figure}
\begin{center}
	\includegraphics[width=0.32\textwidth]{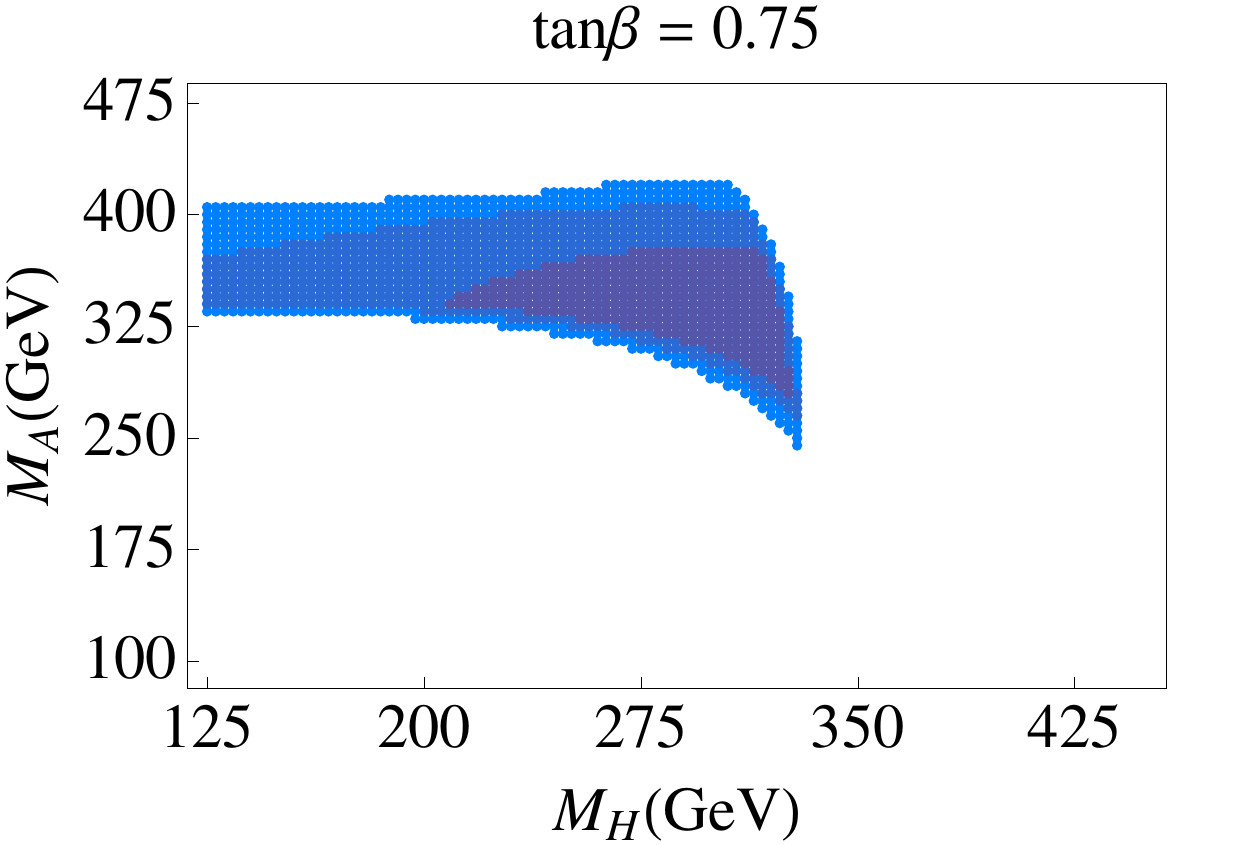}
	\includegraphics[width=0.32\textwidth]{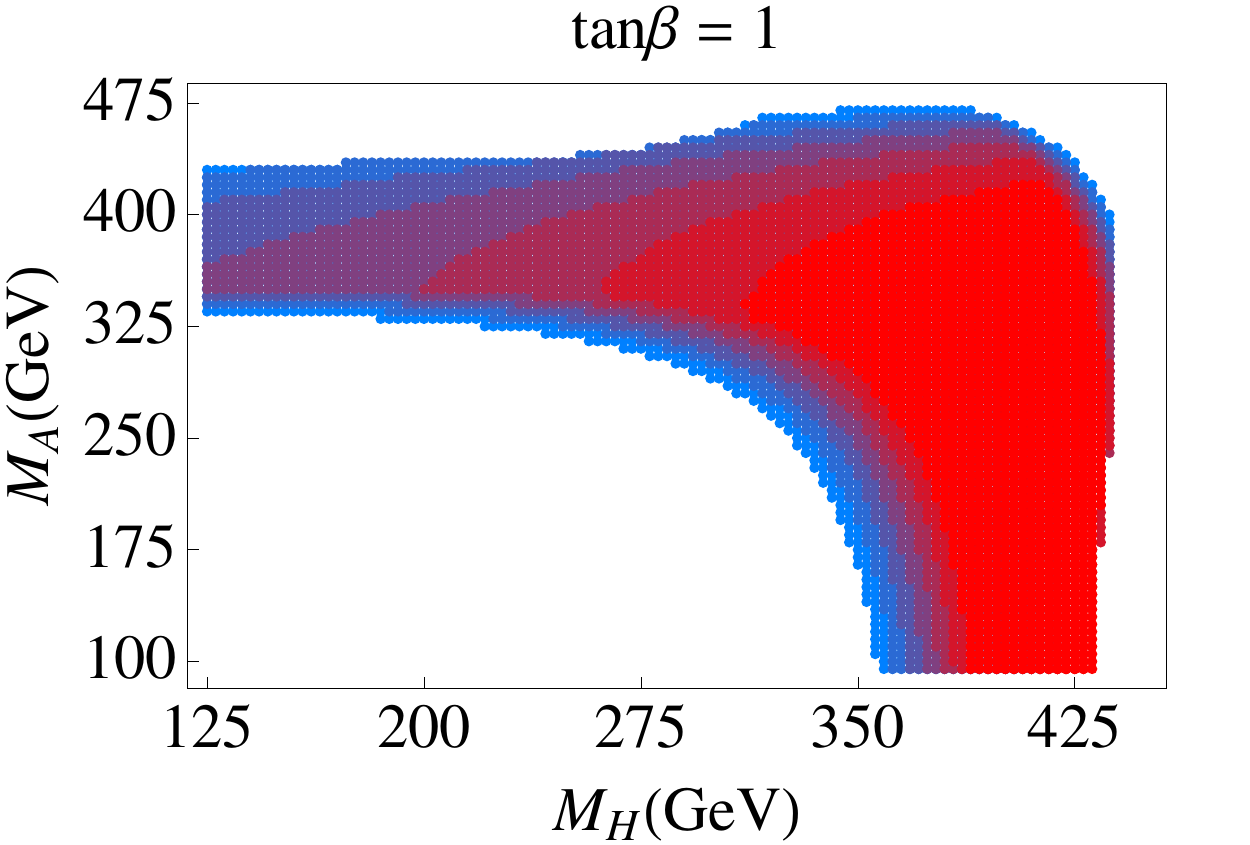}
	\includegraphics[width=0.32\textwidth]{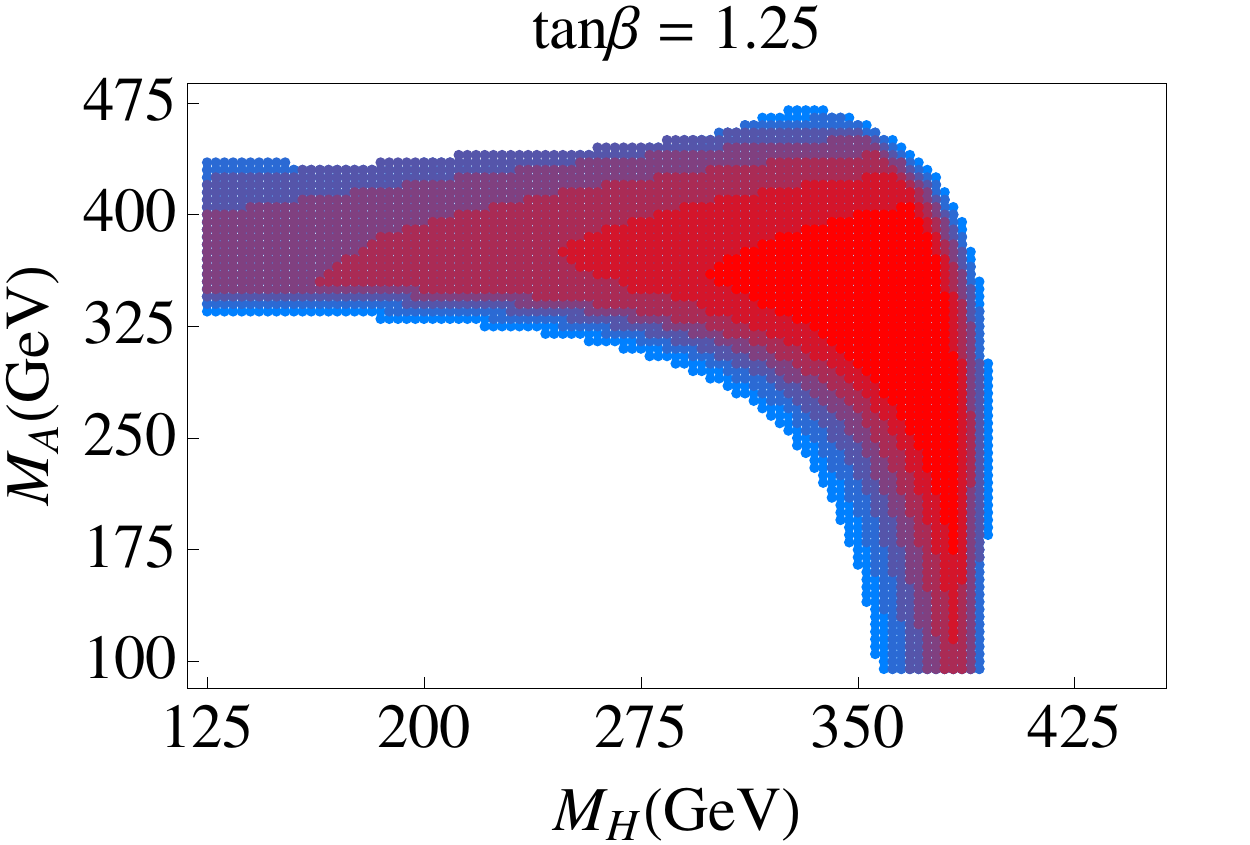}\\[.3cm]
	\includegraphics[width=0.32\textwidth]{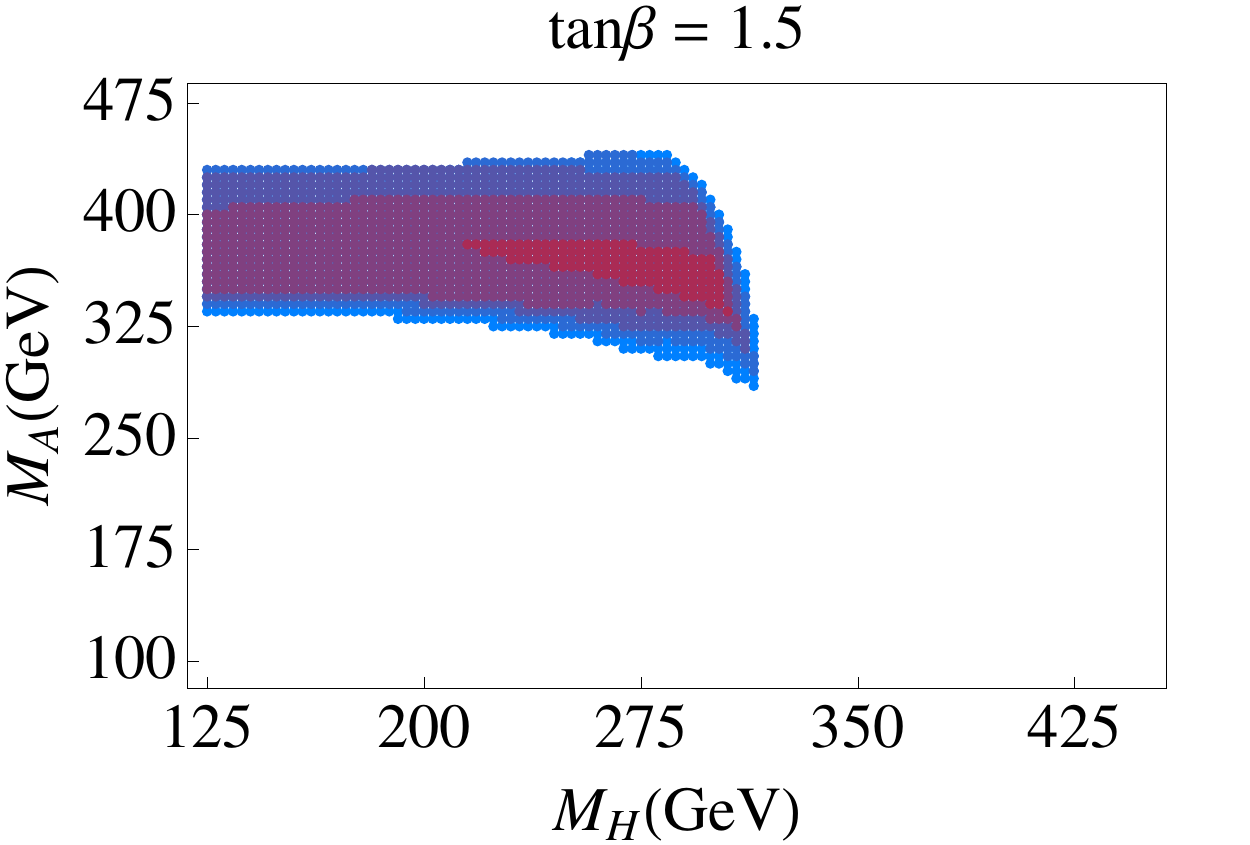}
	\includegraphics[width=0.32\textwidth]{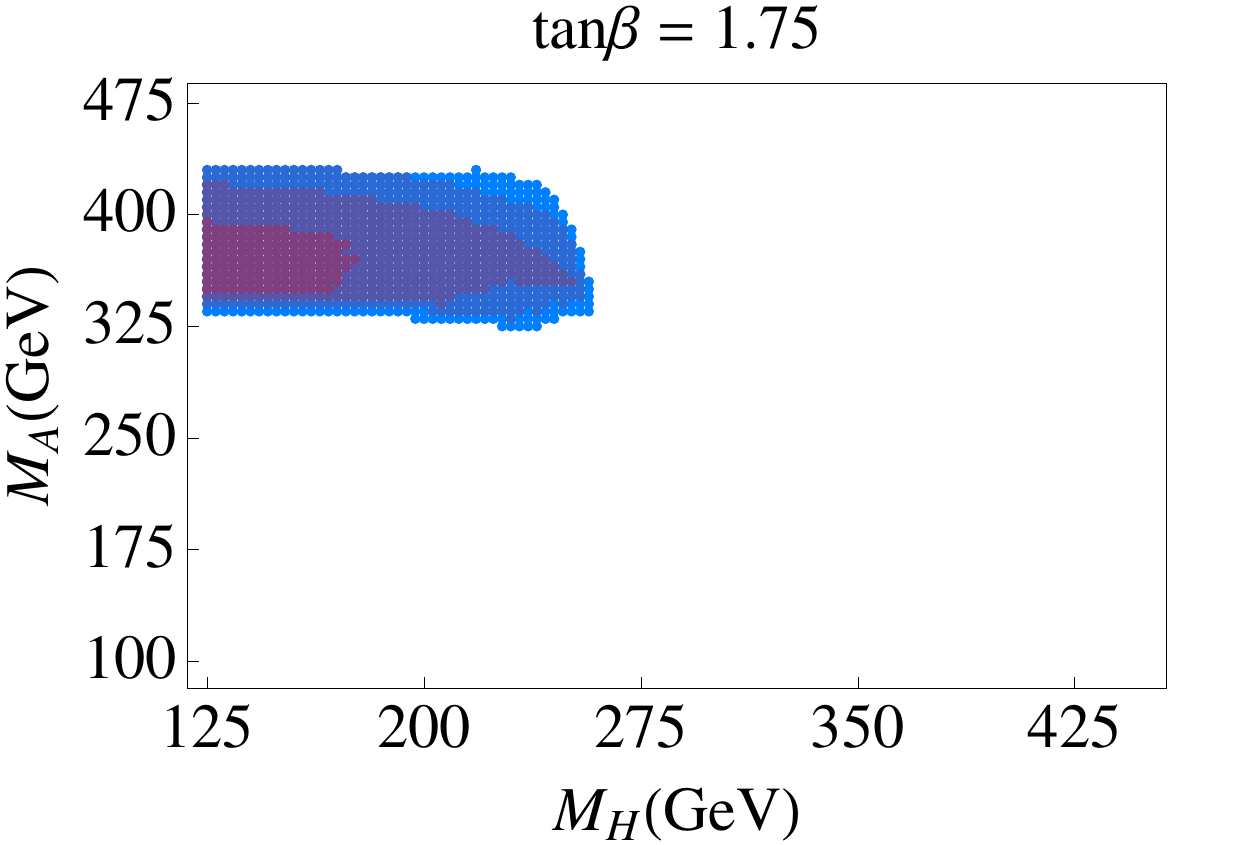}
	\includegraphics[width=0.32\textwidth]{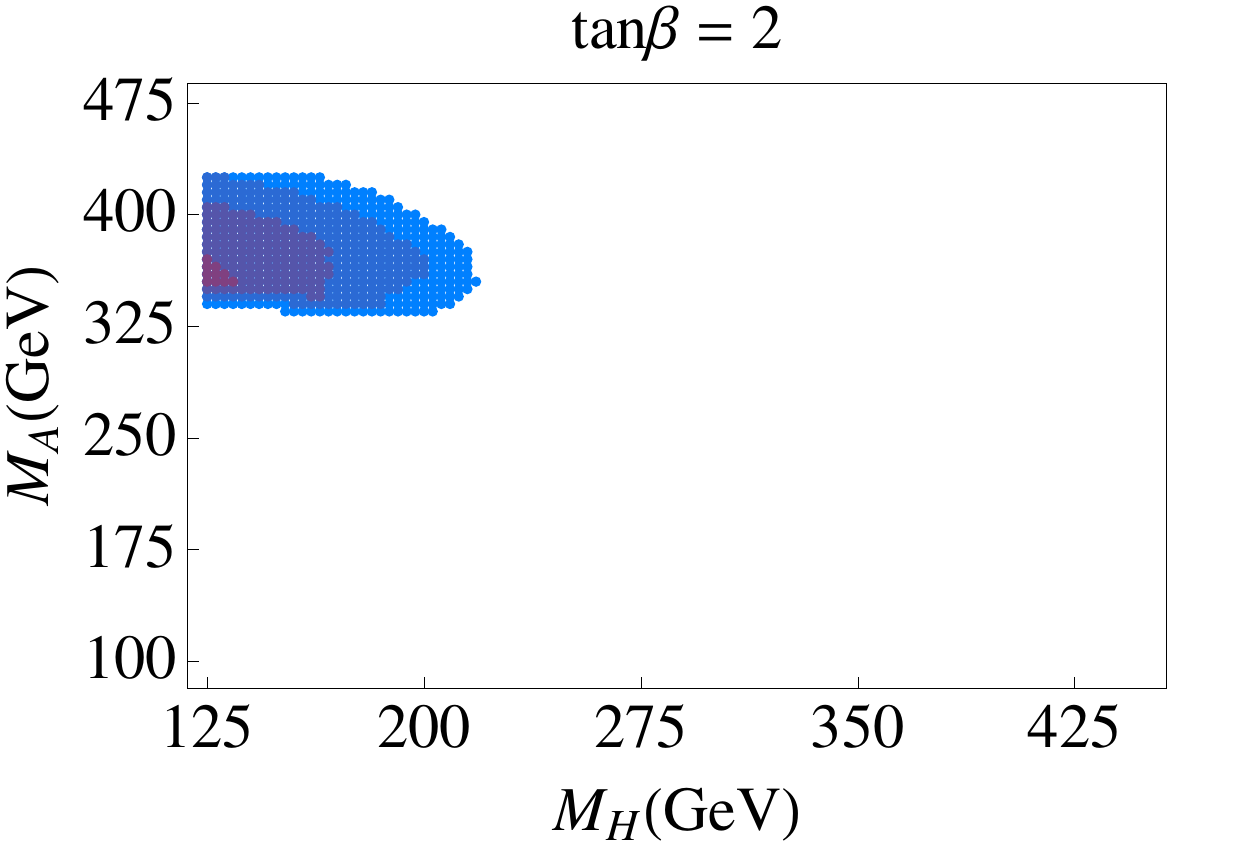}\\[.3cm]
	\includegraphics[width=0.32\textwidth]{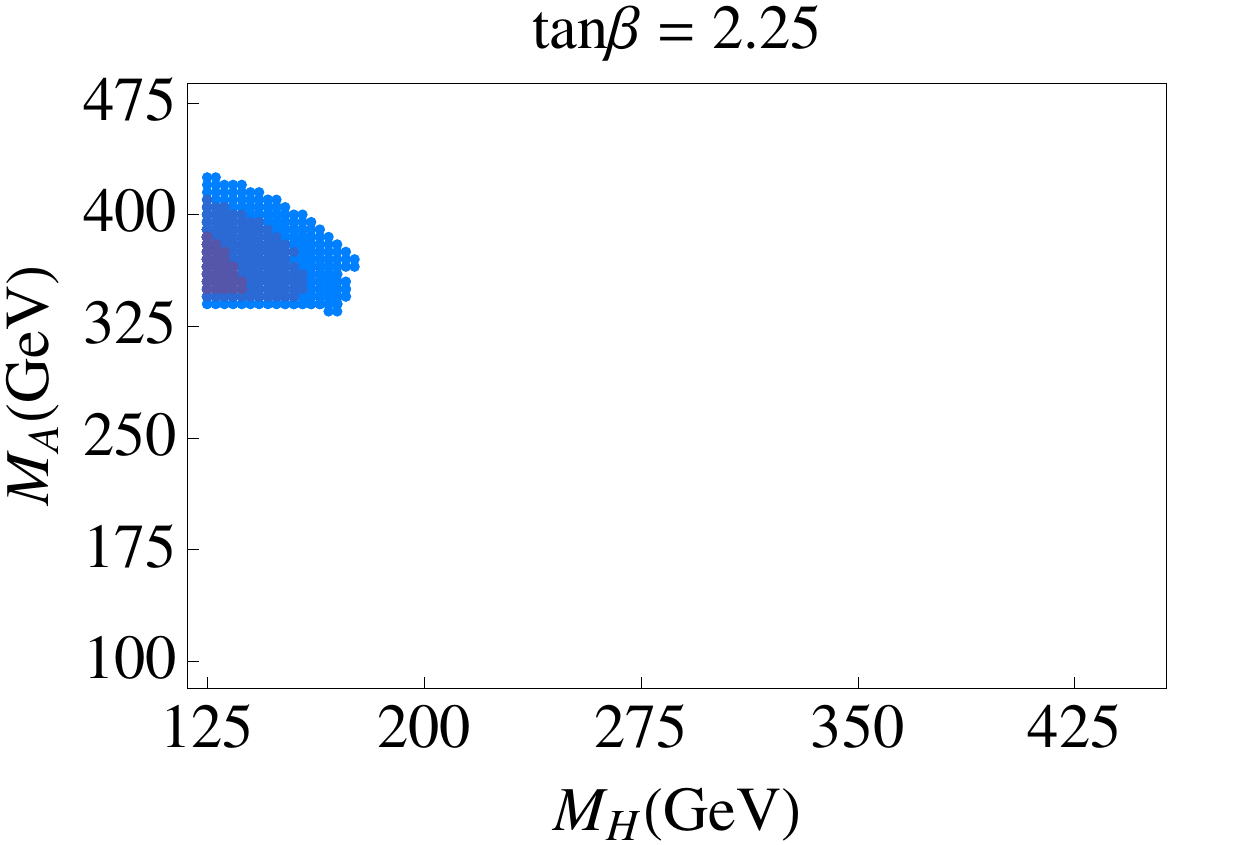}
	\includegraphics[width=0.32\textwidth]{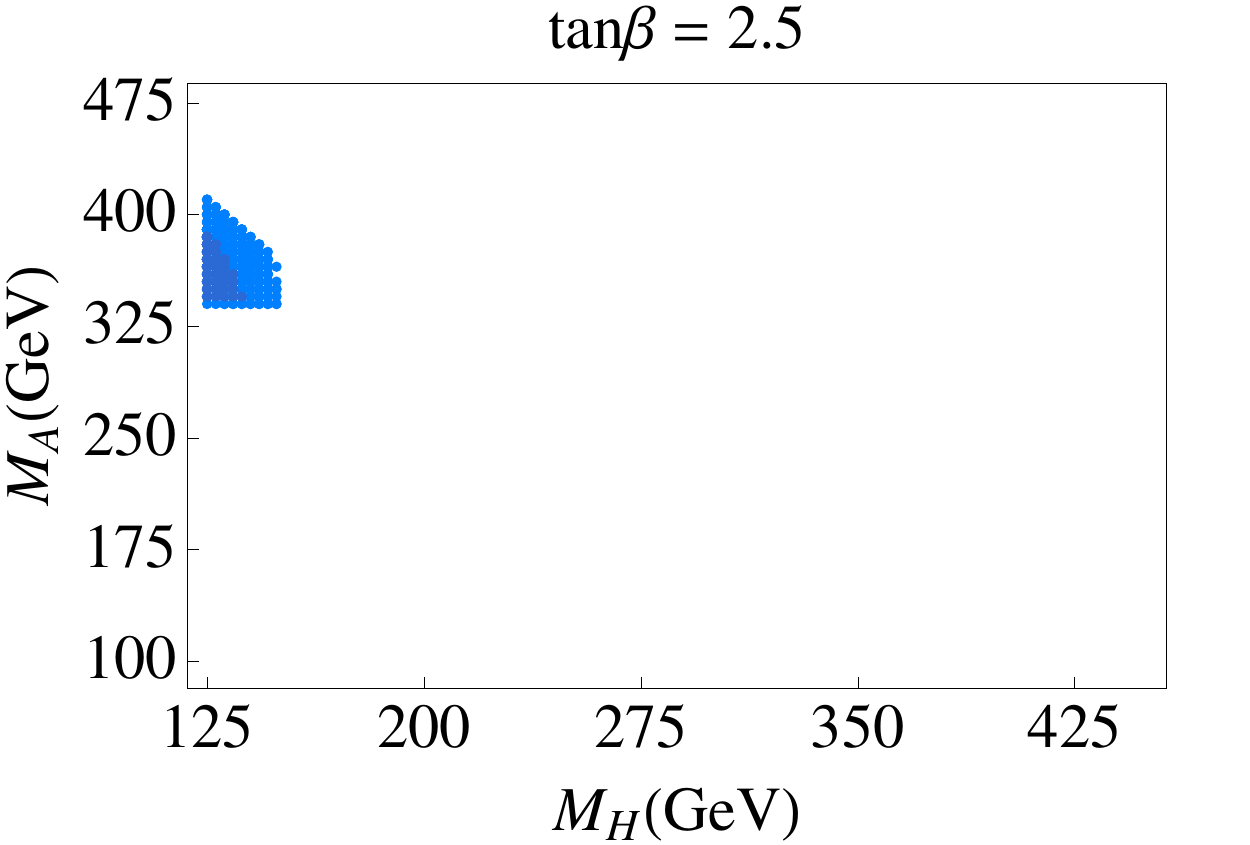}
	\includegraphics[width=0.32\textwidth]{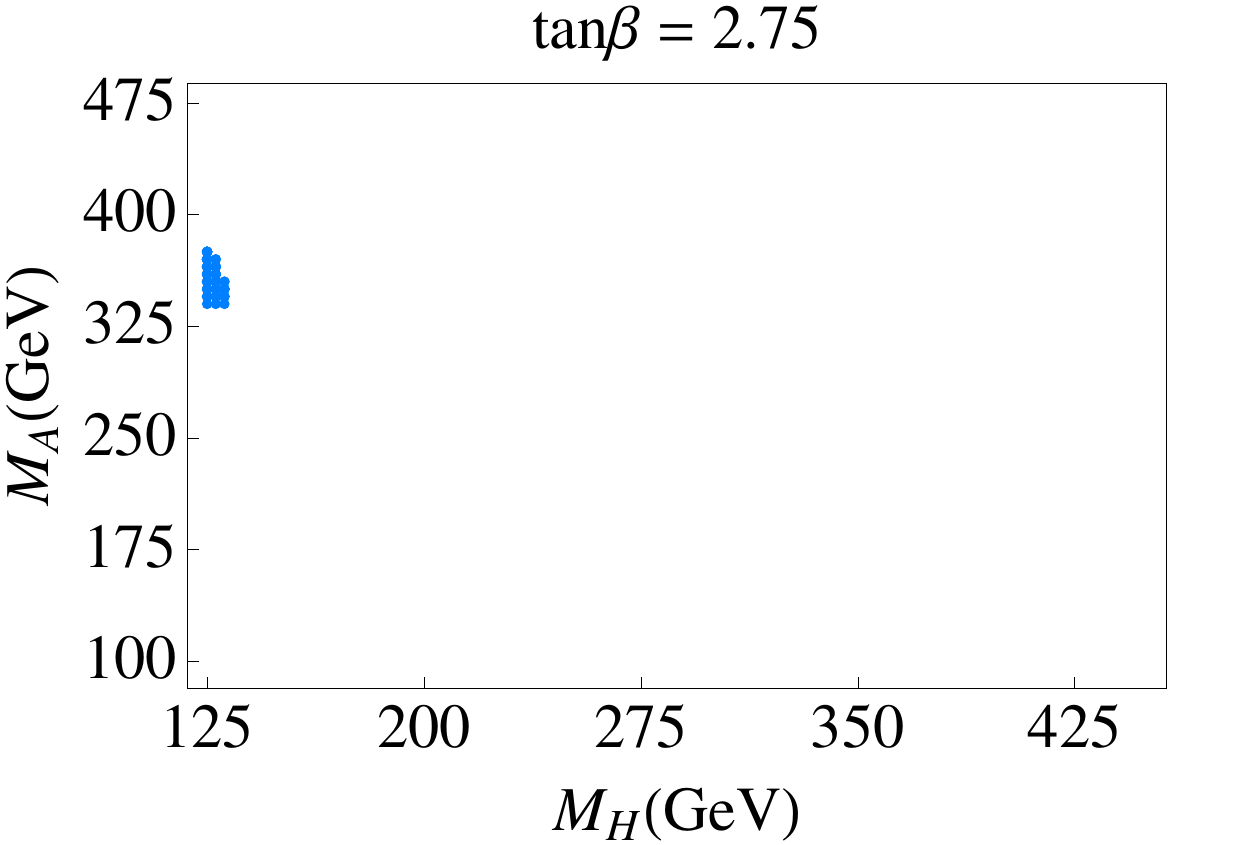}
	\caption{The viable particle spectrum for $0.75\le
          \tan\beta\le2.75$. The spectrum depends non-trivially on the
          charged Higgs mass ($M_{H^\pm}$). In each plot the viable parameter space in $M_H$-$M_A$ plane shrinks as $M_{H^\pm}$ increases. $M_{H^\pm}=380$ (420) GeV corresponds to the color blue (red).}
	\label{fig:viableparam}
\end{center}
\end{figure}

In this section we present the result of our parameter space scan
consistent with both the Higgs data and the experimental and
theoretical constraints discussed above.  
A point in parameter space is  a set of values of the parameters $M_H$, $M_A$, $M_{H^\pm}$, $\alpha$, $\tan\beta$, as discussed in section~\ref{sec:scalarsector}.
For every parameter point consistent with the experimental constraints of~\ref{sec:expconstraints}, we determine the corresponding scalar and Yukawa couplings using equations~\eqref{eq:yukawacouplings} and~\eqref{eq:scalarcouplings}. These couplings are then evolved numerically from the weak scale, $v_w$, to the cut-off scale, $\Lambda$, using beta-functions listed in appendix~\ref{app:beta}. Finally the couplings are checked against the theoretical constraints of section~\ref{subsec:theorybounds}.

The interplay between the
experimental and theoretical bounds on the spectrum can be easily
understood.  Intuitively, experimental bounds--- $Br(\bar{B}\to
X_s\gamma)$ and electroweak precision data tend to drive the mass of
the Higgs bosons heavy in order to leave a small imprint on low energy
observables.  
At the same time, the more massive the
spectrum is, the larger the scalar couplings.  Perturbativity
constraints limit how large these scalar couplings can be, hence limit
from above the Higgs spectrum of the theory\footnote{This conclusion is relaxed somewhat if we allow the $Z_2$ symmetries to be broken softly, see section~\ref{sec:scalarsector}.}.
Combining these experimental and theoretical considerations, we find that
with the cut-off scale ($\Lambda$) at 2 TeV, there is no viable parameter space.
Reducing $\Lambda$ to 1 TeV opens up a small parameter space for small
values of $\tan\beta$. Thus, for the rest of this work we will take $\Lambda=1$ TeV. The viable spectrum is shown in
figure~\ref{fig:viableparam}.

We end this section with a brief discussion of the sensitivity of the viable parameter space on the choice of perturbativity condition, eg. equation~\eqref{eq:perturbativity}.
Had we imposed instead that all the reduced couplings remain less than 1/2, there would be no viable parameter space.  
We find that relaxing the perturbativity constraint from 1/2 to 3/4 opens up a small viable parameter space for the cases $\tan\beta = 1$, 1.25.  
Relaxing this constraint further to 1 leads us to the viable parameter space that we have in figure~\ref{fig:viableparam}.

\section{Phenomenology Of  The Other Higgs Bosons }
\label{sec:pheno}
In this section we study the phenomenology of the neutral CP-even and
CP-odd Higgs bosons, $H$ and $A$.  We will focus on their
production cross-sections and branching ratios.  For a wide range of
production and decay channels, we can deduce the cross-sections and
branching ratios by scaling from the corresponding quantities for the SM
Higgs boson.  When this  scaling procedure is not available, we
compute the corresponding quantity at leading order.

\subsection{Phenomenology of the CP-even $H$}
\begin{figure}
\begin{center}
	\includegraphics[width=0.45\textwidth]{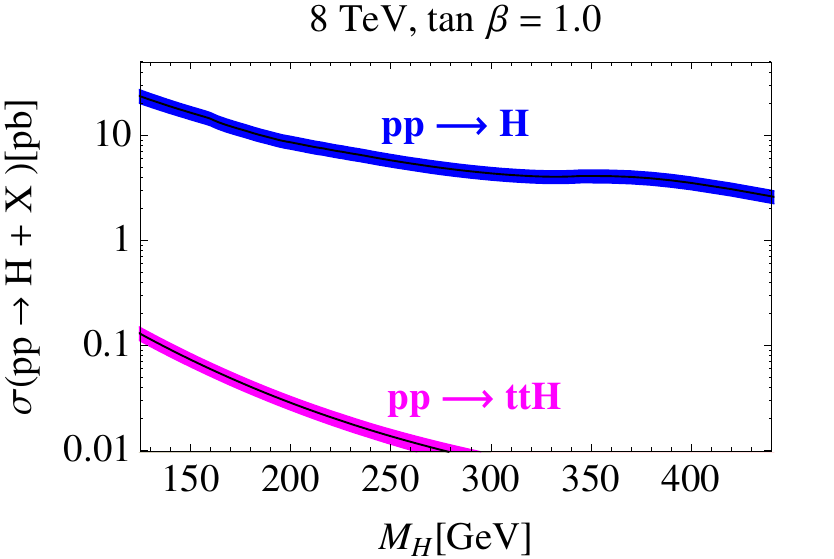}
	\includegraphics[width=0.45\textwidth]{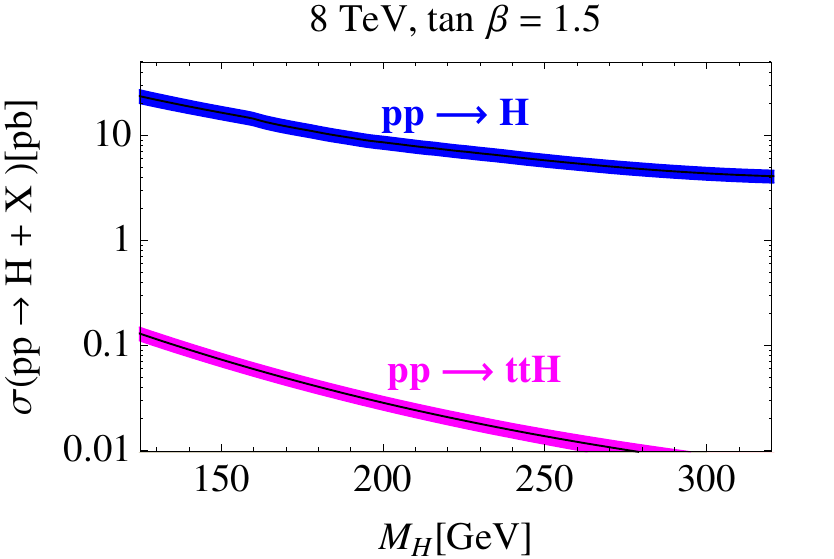}\\[0.2cm]
	\includegraphics[width=0.45\textwidth]{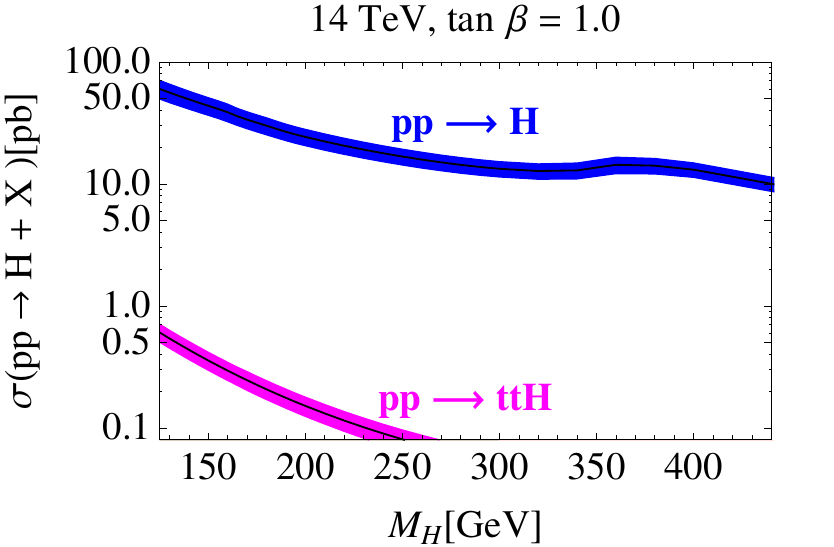}
	\includegraphics[width=0.45\textwidth]{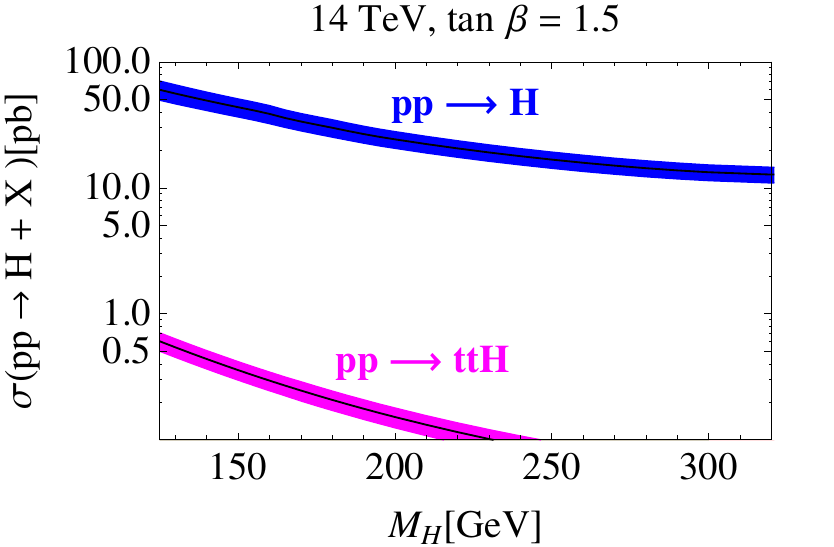}
	\caption{The
          heavy CP-even Higgs boson production cross-sections.  
          For each plot, $\alpha$ is taken to be 0.78 (0.58) for $\tan\beta=1.0\;(1.5)$.
          This choice of $\alpha$ minimizes
          $\chi^2|_{\tan\beta}$ for a given value of $\tan\beta$.}
	\label{fig:Hxsec}
\end{center}
\end{figure}

\begin{figure}
\begin{center}
	\includegraphics[width=0.45\textwidth]{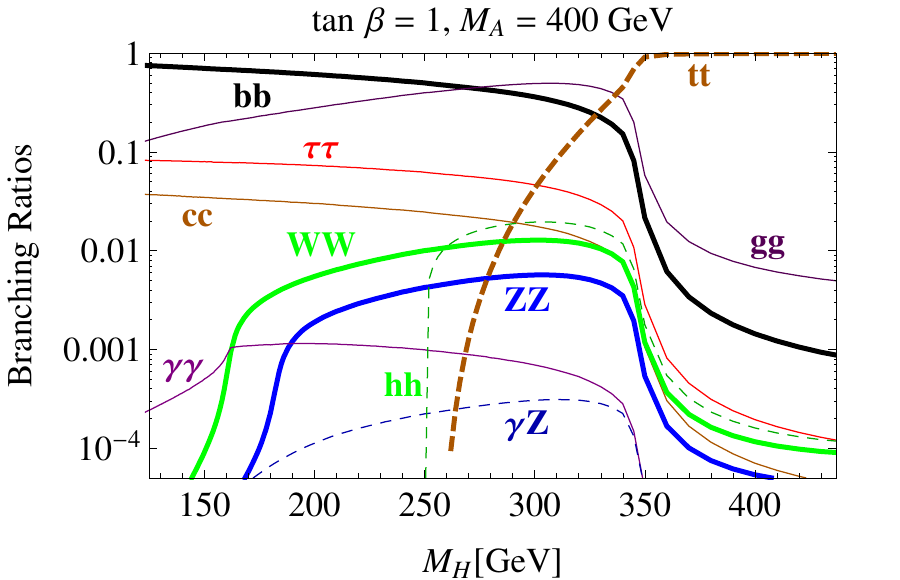}
	\includegraphics[width=0.45\textwidth]{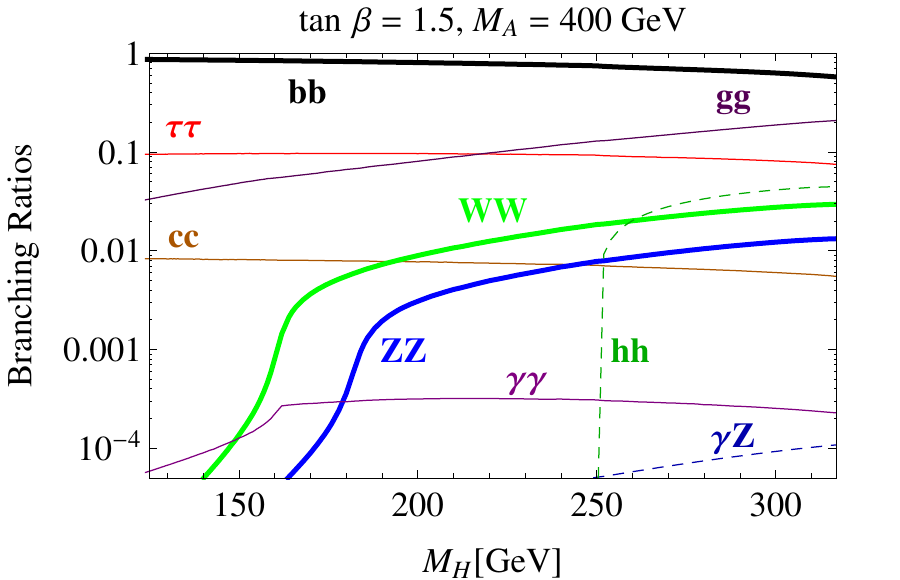}
	\caption{The heavy CP-even Higgs boson branching ratios for the case of heavy $M_A$. 
	For each plot, $\alpha$ is taken to be 0.78 (0.58) for $\tan\beta=1.0\;(1.5)$.
	This choice of $\alpha$ minimizes
          $\chi^2|_{\tan\beta}$ for a given value of $\tan\beta$.}
	\label{fig:Hbr}
\end{center}
\end{figure}

\begin{figure}
\begin{center}
	\includegraphics[width=0.7\textwidth]{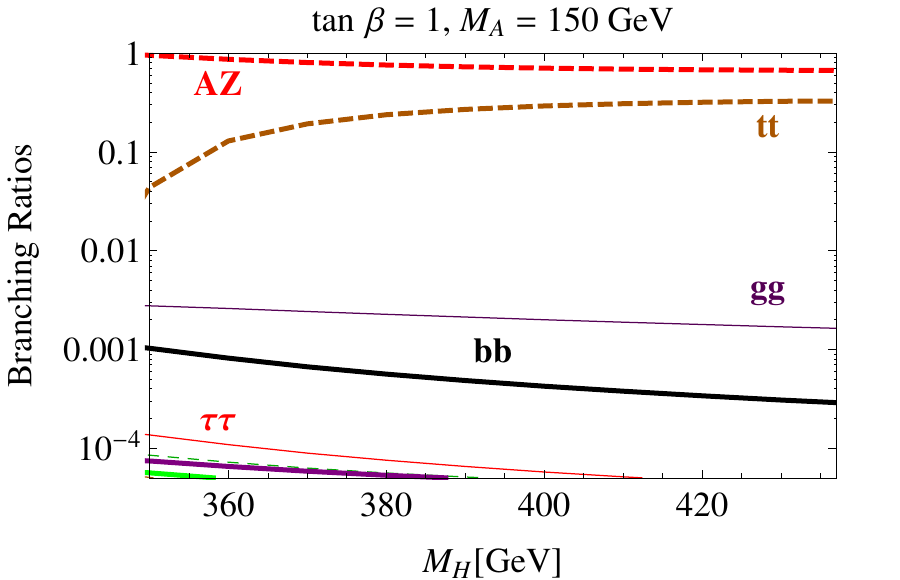}
	\caption{The heavy CP-even Higgs boson branching ratios for the case of light $M_A$. 
	Here $\alpha$ is taken to be 0.78.
	This choice of $\alpha$ minimizes
          $\chi^2|_{\tan\beta}$ for a given value of $\tan\beta$.}
	\label{fig:HbrwithA}
\end{center}
\end{figure}

The couplings of the heavy CP-even Higgs boson $H$ to the SM fermions
and gauge bosons are modified with respect to the SM Higgs boson
couplings by the rescaling factors
\begin{equation}
\begin{split}
	c^H_t = c^H_c = \frac{-\sin\alpha}{\sin\beta},\quad
	c^H_b = c^H_\tau = \frac{\cos\alpha}{\cos\beta},\quad
	c^H_V = \cos(\alpha+\beta).
\end{split}
\end{equation}
The $H$ production cross-sections can be readily obtained by rescaling
the corresponding calculations for SM Higgs boson production.  The
CP-even $H$ boson production cross-sections for the 8 and 14 TeV LHC
are shown in figure~\ref{fig:Hxsec}.  Note that since the mixing angles $\alpha$ and $\beta$ are close to the decoupling limit, the $H$ production cross-sections in the vector boson fusion and the $W$- and $Z$-associated production channels are suppressed.

In a large portion of the available parameter space $M_H$ is lighter
than $2M_h$, $M_A$ and $M_{H^\pm}$; see figure~\ref{fig:viableparam}.
When this is the  case all the available decay channels for the heavy Higgs
boson, $H$, are the same as for the SM Higgs boson.
The branching ratios can be obtained by a simple rescaling from the
corresponding values of the SM Higgs boson.
When $M_H$ is sufficiently heavy, the decays $H\to hh$, $H\to AZ$ and
$H\to AA$ become available.  The partial widths are\footnote{Our result
  does not agree with that of Ref~\cite{Djouadi:2005gj} whose partial
  width for $\Phi\to\phi Z$ does not have a correct mass dimension.}
\begin{equation}
\begin{split}
	\Gamma(H\to \phi\phi) &= \frac{G_F}{8\sqrt{2}\pi}\frac{\bar{\lambda}_{H\phi\phi}^2}{M_H}\sqrt{1-\frac{4M_\phi^2}{M_H^2}}\\
	\Gamma(H\to AZ) &= \frac{G_F}{8\sqrt{2}\pi}\sin^2(\alpha+\beta)\left[\frac{\left(M_H^2-(M_Z-M_A)^2\right)\left(M_H^2-(M_Z+M_A)^2\right)}{M_H^2}\right]^{3/2},
\end{split}
\end{equation}
where $G_F$ is the Fermi constant, $\phi = h$, $A$ and
\begin{equation}
\begin{split}
	\bar{\lambda}_{Hhh} &= \frac{\left(2M_h^2+M_H^2\right)\cos(\alpha+\beta)\sin(2\alpha)}{2\sin\beta\cos\beta},\\
	\bar{\lambda}_{HAA} &= \frac{\left(3M_H^2+2M_A^2\right)\sin(\alpha-\beta)+\left(M_H^2-2M_A^2\right)\sin(\alpha+3\beta)}{4\sin\beta\cos\beta}.
\end{split}
\end{equation}

The branching fractions for $H$ decays are shown in
  figure~\ref{fig:Hbr} and~\ref{fig:HbrwithA}.  Since the mixing
  angles $\alpha$ and $\beta$ are close to the decoupling limit, the
  decays into a pair of  massive vector bosons are
  suppressed.   When $H\to A+X$ is kinematically
    forbidden, the $H$ boson decays predominantly into quarks and
    gluons; see figure~\ref{fig:Hbr}.  In this case, 
    it is most likely to observe the $H$-boson in di-photon decays.
  However, when $H\to A+X$ is kinematically allowed, the
    decay mode $H\to AZ$ becomes the most dominant; see
    figure~\ref{fig:HbrwithA}.  As we will show in the next section,
    the $A$ has sizable branching ratios into $\gamma Z$ and
    $\gamma\gamma$.  Thus it might be possible to observe $H\to AZ$ in
    a photon plus four leptons or two photons and two leptons
    channels.

\subsection{Phenomenology of the CP-odd Boson}
\begin{figure}
\begin{center}
	\includegraphics[width=0.45\textwidth]{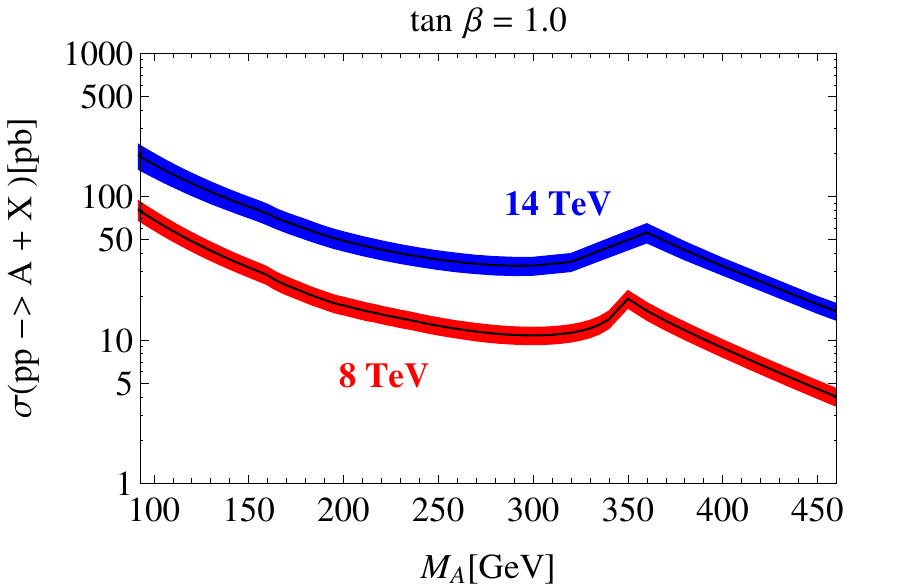}
	\includegraphics[width=0.45\textwidth]{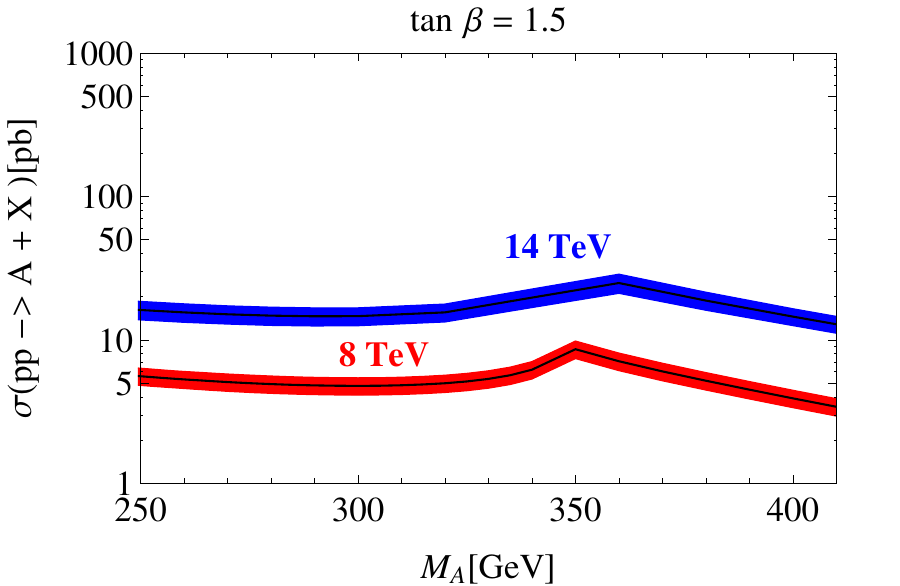}
	\caption{The heavy CP-odd Higgs boson production cross-sections. 
          For each plot, $\alpha$ is taken to be 0.78 (0.58) for $\tan\beta=1.0\;(1.5)$. 
          This choice of $\alpha$ minimizes
          $\chi^2|_{\tan\beta}$ for a given value of $\tan\beta$.}
	\label{fig:Axsec}
\end{center}
\end{figure}

The pseudoscalar $A$ does not couple at tree level to a pair of
electroweak gauge bosons.  Thus its main production mechanism is via
gluon fusion.  Its production cross-section can be
obtained from that of the Higgs in the Standard Model by  rescaling by a  factor
\begin{equation}
	r_g \simeq \left|\cot\beta\frac{\tau f(\tau)}{\tau+(\tau-1)f(\tau)}\right|^2,
\end{equation} 
where $\tau = M_A^2/4M_t^2$ and the function $f(\tau)$ is defined in appendix~\ref{app:decays}.
The production cross-sections of the pseudoscalar $A$ at the LHC with
8 TeV and 14 TeV center-of-mass energy are shown in
figure~\ref{fig:Axsec}.

\begin{figure}
\begin{center}
	\includegraphics[width=0.45\textwidth]{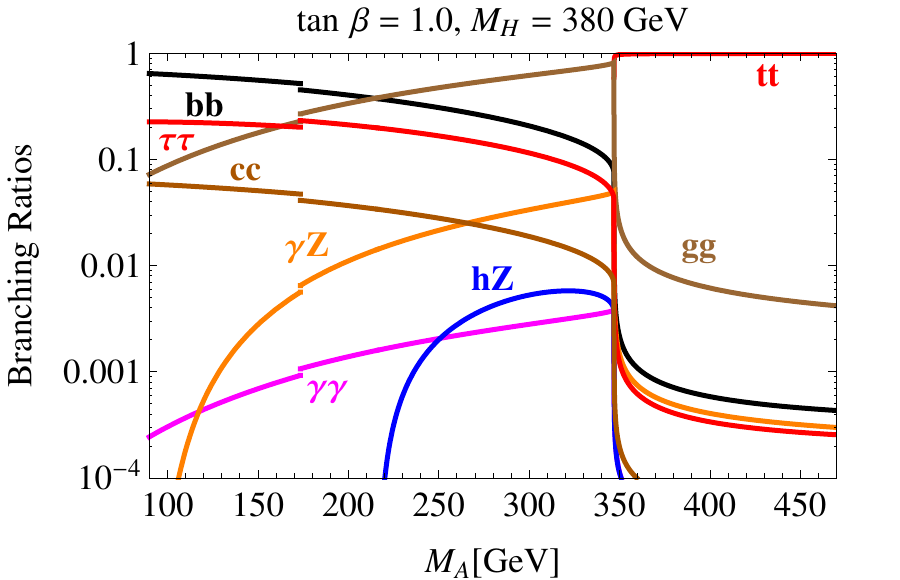}
	\includegraphics[width=0.45\textwidth]{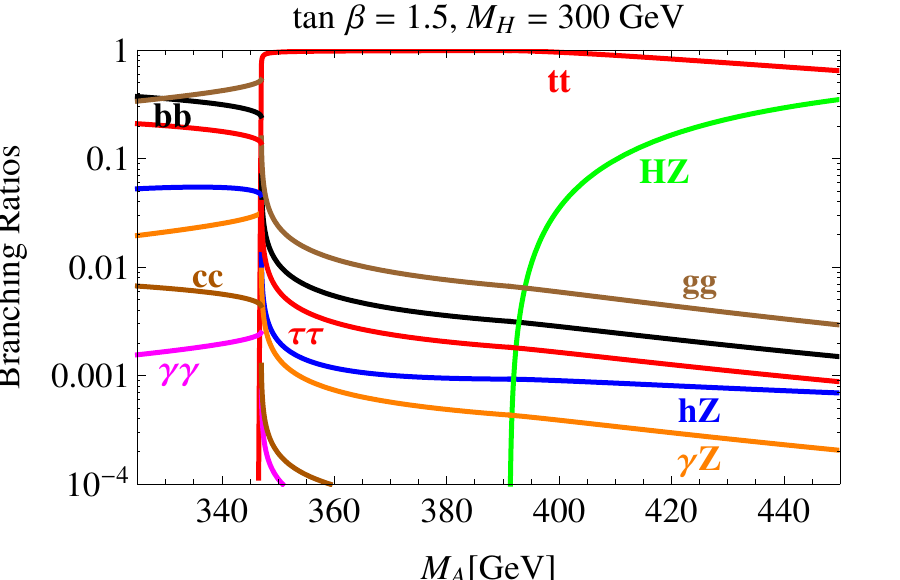}\\[.2cm]
	\includegraphics[width=0.45\textwidth]{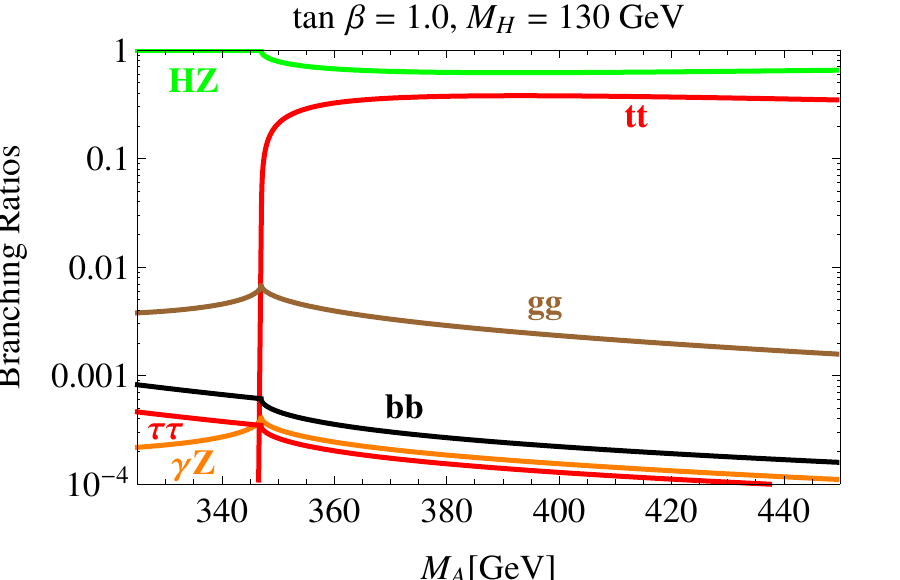}
	\includegraphics[width=0.45\textwidth]{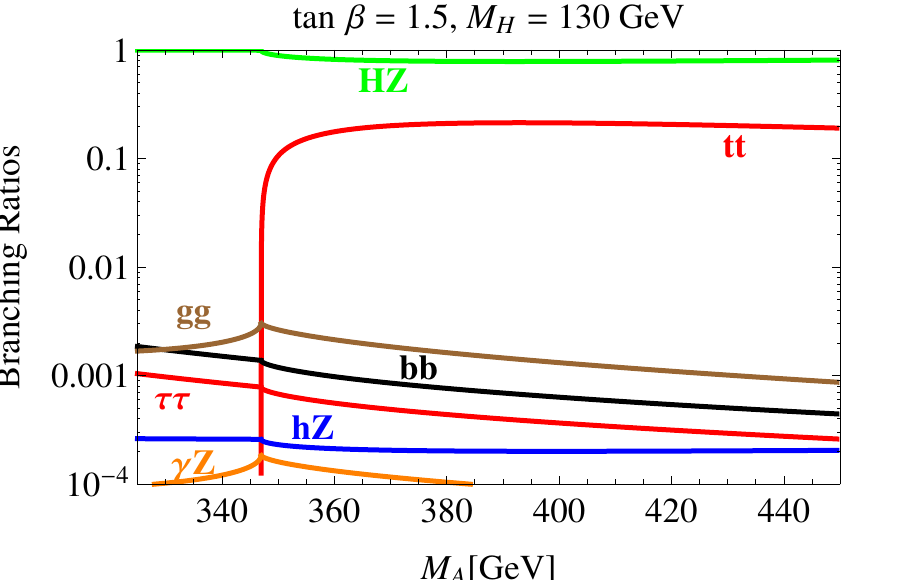}
	\caption{The CP-odd Higgs boson branching ratios.
	 For each plot, $\alpha$ is taken to be 0.78 (0.58) for $\tan\beta=1.0\;(1.5)$.
	 This choice of $\alpha$ minimizes
          $\chi^2|_{\tan\beta}$ for a given value of $\tan\beta$. Diagrams on the top correspond to the case when $M_H$ is sufficiently heavy and cannot be in the decay product of $A$.  Diagrams on the bottom are when $M_H$ is light enough to be in the decay product. For definiteness, we take $M_H=130$ GeV. Here we ignore the effect of near threshold production}
	\label{fig:Abr}
\end{center}
\end{figure}

The CP-odd boson $A$ has a mass $M_A \lesssim
470$ GeV, see figure~\ref{fig:viableparam}.  A light $A$ can decay
into a pair of light quark-antiquark as well as into two photons, much
like the SM Higgs.  For sufficiently large $M_A$, $A$ can also decay
into pair consisting of a neutral CP-even scalar boson and a vector
boson, $\phi Z$, where $\phi=h,H$. 
The partial decay widths for $A\to q\bar{q}$ is
\begin{equation}
	\Gamma(A\to q\bar{q}) = 3\frac{G_F M_q^2}{4\sqrt{2}\pi} \delta_q^2 M_A \sqrt{1-4M_q^2/M_A^2},
\end{equation}
where $\delta_q = \cot\beta$ $(\tan\beta)$ for up-type (down-type) quark. 
The partial decay width into $\phi Z$ is given by
\begin{equation}
\label{eq:atophiz}
	\Gamma(A\to \phi Z) = \frac{G_F}{8\sqrt{2}\pi}\delta_\phi^2\left[\frac{\left(M_A^2-(M_Z-M_\phi)^2\right)\left(M_A^2-(M_Z+M_\phi)^2\right)}{M_A^2}\right]^{3/2},
\end{equation}
where $\delta_h$ ($\delta_H$) = $\cos(\alpha+\beta)$
($\sin(\alpha+\beta)$).  
The loop-induced branching ratios into $gg$, $\gamma\gamma$ and $\gamma Z$ are given in appendix~\ref{app:decays}.
The CP-odd $A$ branching ratios are shown in
figure~\ref{fig:Abr}. Note that the suppression of $Br(A\to hZ)$ 
arises because the two mixing angles $\alpha$ and $\beta$ are close to the decoupling limit, see equation~\eqref{eq:atophiz}.  
The dominant decay channels for $A$ are $b\bar{b}$ and $gg$ for a low
mass $A$ and $HZ$ and $t\bar{t}$  for a heavy
$A$. 
  
\begin{figure}
\begin{center}
	\includegraphics[width=0.75\textwidth]{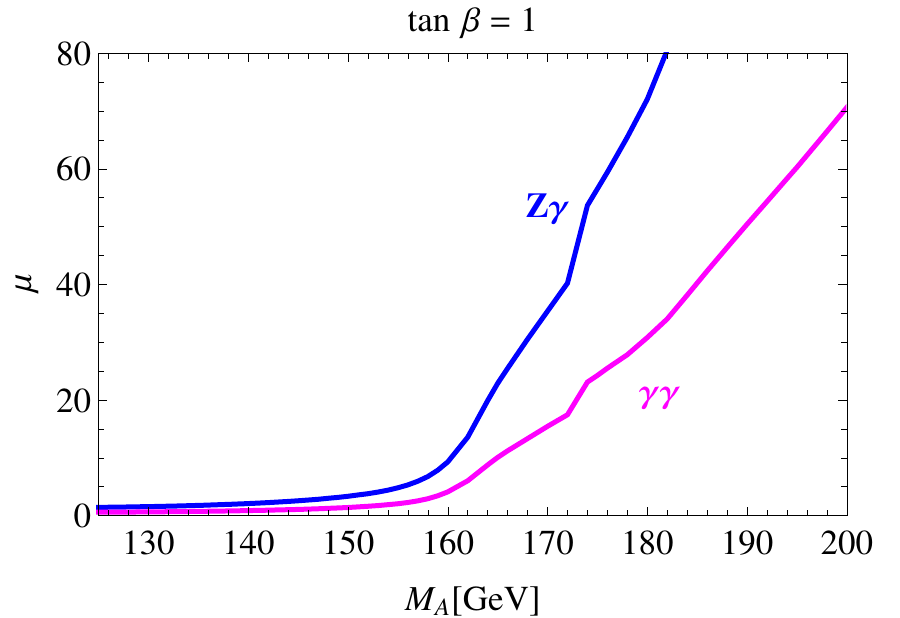}
	\caption{The signal strength $\mu(A\to\gamma \gamma)$ and $\mu(A\to\gamma Z)$ for the case $\tan\beta=1$ in the low mass range.}
	\label{fig:muA}
\end{center}
\end{figure}

 It is also interesting to note that for the low mass
  range below the $t\bar{t}$ and $HZ$ thresholds, the $A$ production
  cross-section and branching ratios into $\gamma \gamma$ and $\gamma
  Z$ are enhanced compared to the corresponding Standard Model Higgs
  boson counterparts.  Our estimate of the signal strength in these
  two modes is shown in figure~\ref{fig:muA}.  With the large signal
  strength for $M_A\ge160$ GeV, this scenario could be excluded using
  current Higgs data.  However, at the moment both the ATLAS and CMS
  collaborations only provide the 95\% exclusion limit on the signal
  strength in these channels up to a mass of 150
  GeV~\cite{ATLAS-CONF-2013-009,ATLAS-CONF-2013-012,CMS-PAS-HIG-13-006,CMS-PAS-HIG-13-001}.

\subsection{Phenomenology of the Charged Boson $H^{\pm}$}
\begin{figure}
\begin{center}
	\includegraphics[width=0.48\textwidth]{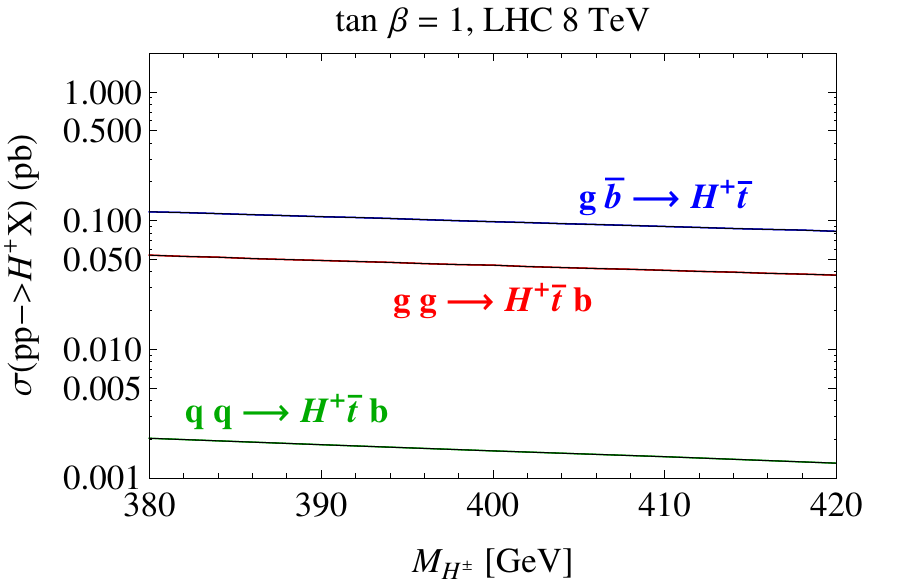}
	\includegraphics[width=0.48\textwidth]{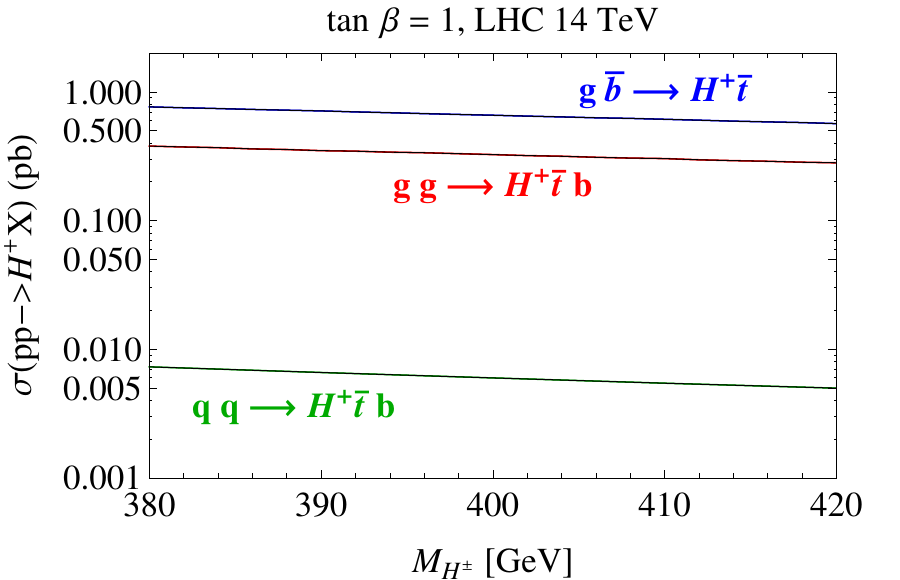}\\[.2cm]
	\includegraphics[width=0.48\textwidth]{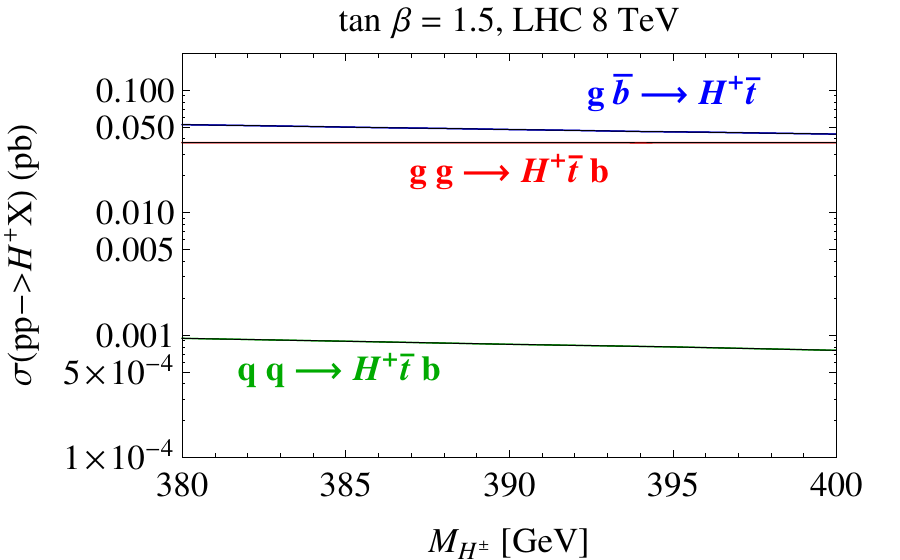}
	\includegraphics[width=0.48\textwidth]{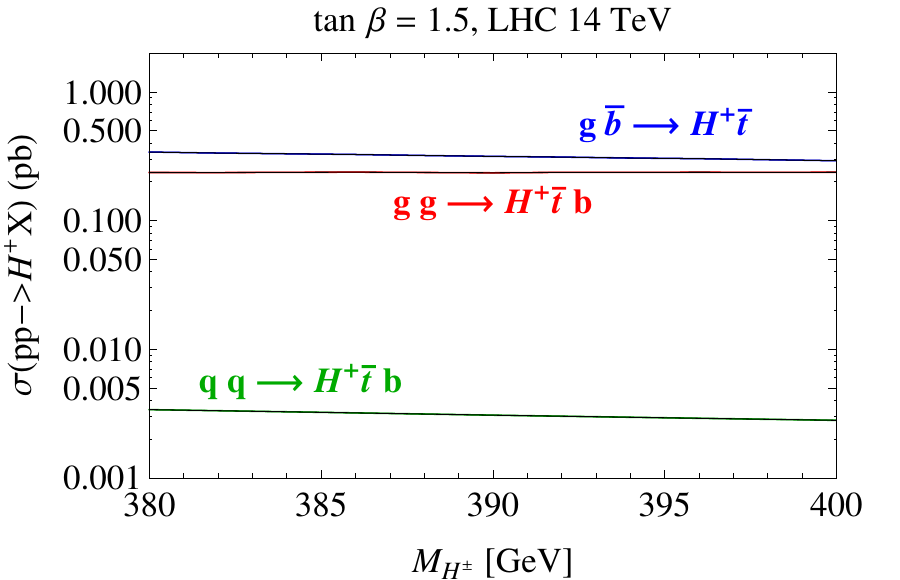}
	\caption{The charged Higgs boson production cross-sections for the case $\tan\beta=1.0\;(1.5)$ for the left (right) plot with $\alpha = 0.78\;(0.58)$.  $\alpha$ is chosen such that it minimizes $\chi^2|_{\tan\beta}$ for a given value of $\tan\beta$. Diagrams on the top correspond to the 8 TeV LHC while the bottom are for 14 TeV.}
	\label{fig:Hpxsec}
\end{center}
\end{figure}

The viable mass of the charged Higgs boson is larger than that of the
top-quark.  Thus its main production cross-section is from
$g\bar{b}\to H^+\bar{t}$ and $gg\to H^+\bar{t}b$~\cite{Gunion:1989we}.
We use Madgraph to compute the lowest order production cross-sections
for both processes.  We follow Ref.~\cite{Borzumati:1999th} in keeping
the renormalization scale and the PDF factorization scale fixed at
$\mu = M_t + M_{H^\pm}$.  For comparison, we also compute the
production cross-section from $q\bar{q}\to H^+\bar{t}b$ where $q$ is
the valence quarks.  The production cross-sections are shown in
figure~\ref{fig:Hpxsec}.

\begin{figure}
\begin{center}
	\includegraphics[width=0.45\textwidth]{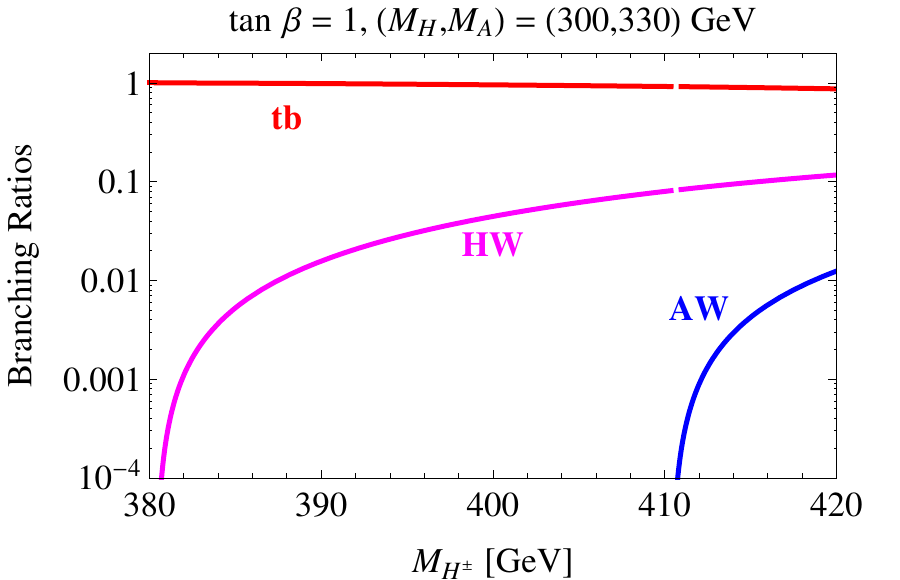}
	\includegraphics[width=0.45\textwidth]{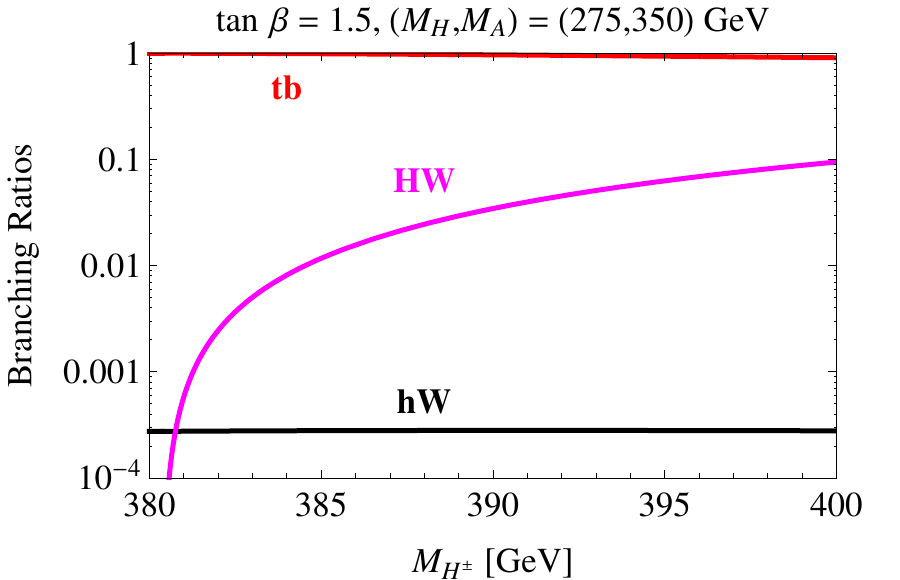}\\[.2cm]
	\includegraphics[width=0.45\textwidth]{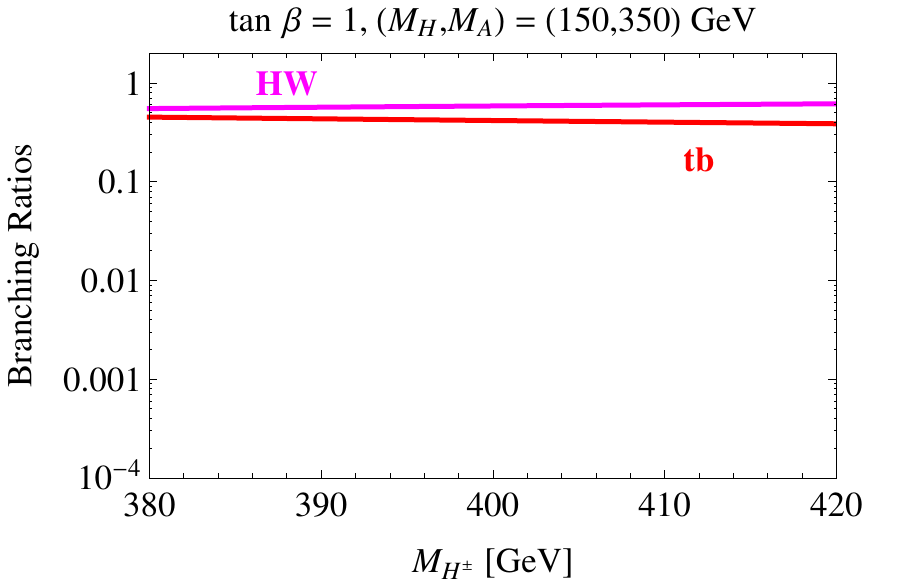}
	\includegraphics[width=0.45\textwidth]{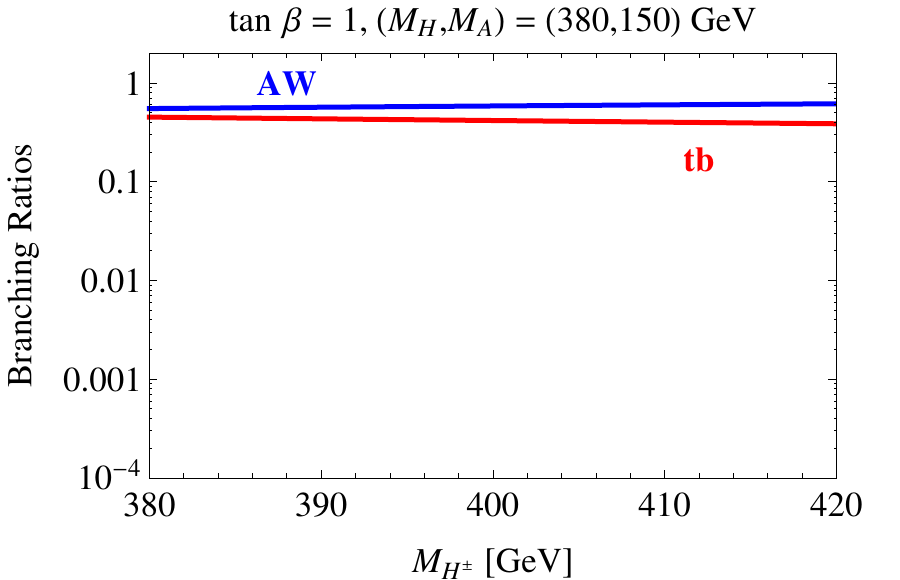}
	\caption{The charged Higgs boson branching ratios for the case
          $\tan\beta=1.0\;(1.5)$ for the left (right) plot with
          $\alpha = 0.78\;(0.58)$.  $\alpha$ is chosen such that it
          minimizes $\chi^2|_{\tan\beta}$ for a given value of
          $\tan\beta$. The masses of the neutral Higgs boson are
          chosen so that they are consistent with the viable parameter space shown in figure~\ref{fig:viableparam}}
	\label{fig:Hpbr}
\end{center}
\end{figure}

The physical charged Higgs couplings to quarks are given by
\begin{equation}
	(\sqrt{2}G_F)^{1/2}\left[\cot\beta (m_u)_i\bar{u}_i V^\ast_{ij} P_L d_j  +\tan\beta (m_d)_j\bar{u}_i V^\ast_{ij} P_R d_j\right]H^+ + \hc,
\end{equation}
where $m_u$ ($m_d$) is the mass matrix of the up (down) type quarks
and $V$ is the CKM matrix. Thus its partial width $\Gamma(H^+\to
t\bar{b})$ at leading order is
\begin{equation}
\begin{split}
	\Gamma(H^+\to t\bar{b}) &= \frac{3G_F}{8\sqrt{2}\pi}|V_{tb}|^2\sqrt{\frac{(M_{H^\pm}^2-(M_t^2+M_b^2))(M_{H^\pm}^2-(M_t^2-M_b^2))}{M_{H^\pm}^2}}\\
	&\qquad \left[\left(1-\frac{M_t^2}{M_{H^\pm}^2}-\frac{M_b^2}{M_{H^\pm}^2}\right)\left[M_t^2\cot^2\beta + M_b^2\tan^2\beta\right]+\frac{4M_t^2M_b^2}{M_{H^\pm}^2}\right].
\end{split}
\end{equation}
The charged Higgs can also decay into a neutral scalar and a vector.
In this case the partial width can be read off from
equation~\eqref{eq:atophiz} with the replacement $M_A\to M_{H^\pm}$
and $\delta_A = 1$.  The branching ratios of the charged Higgs boson
are shown in figure~\ref{fig:Hpbr}.  Much
    like for $A\to hZ$, the branching ratio for  $H^+\to hW^+$ is
  suppressed because $\alpha$ and $\beta$ are close to the decoupling
  limit. For a light $H(A)$, the decay $H^\pm\to H(A)W^\pm$ is
  dominant.  This decay leads to an event with 3 or 4 $b$-quarks
  (depending on the production channel) plus a charged lepton and
  missing energy which could be searched for at the LHC.

\section{Conclusions and Discussions}
\label{sec:conclusion}
We have performed a parameter scan for the CP conserving 2HDM-II consistent with all
the available Higgs data.  We take into account theoretical bounds ---
vacuum stability and perturbativity of the couplings, as well as
experimental bounds from electroweak precision measurements and
$Br(\bar{B}\to X_s\gamma$) in our scan.  We use a working assumption
that the 2HDM-II is a valid low energy effective theory up to a
cut-off scale $\Lambda$.  We found that for $\Lambda \ge 2$ TeV, there
is no viable parameter space consistent with all the mentioned
constraints.  However, if we assume $\Lambda=1$ TeV, only a small
parameter space of the 2HDM-II is consistent with all the mentioned
constraints.  In particular, the ratio of the vacuum expectation value
of the two scalar doublets, $\tan\beta$, lies in the range
$0.75\lesssim \tan\beta \lesssim 2.75$.

Our results show that the charged Higgs boson is the most constrained
sector of the 2HDM-II.  The perturbativity constraint demands
$M_{H^\pm}\le 420$ GeV while constraint from $Br(\bar{B}\to
X_s\gamma)$ pushes $M_{H^\pm}\ge380$ GeV.  Since the mass of the
charged Higgs boson is much heavier than that of the top-quark, its
main production mechanism is production in association with the top or
in association in the top and the bottom-quarks.  The
  charged Higgs, once produced, decays mostly into a top- and
  bottom-quark pair which makes its detection difficult.  However, in
  a corner of parameter space where the heavy CP-even $H$ (or the
  CP-odd $A$) is light enough, the decay into $H(A)W$ becomes
  comparable to the $tb$ channel.  The light $H(A)$ then  mostly
  decays into a  $b\bar{b}$ pair.  This decay chain leads to an
  event with 3 $b$-quarks (or 4, depending on the production mechanism)
  plus a charged lepton and missing energy which makes  it possible to
  be searched for at the LHC.  

The neutral scalars sector is not as tightly constrained.  The masses
of both the heavy CP-even Higgs boson, $H$, and the CP-odd, $A$, can
take on values in a large range; see figure~\ref{fig:viableparam}.
The production cross-sections for these two particle are sizable at
both the 8 TeV and the 14 TeV LHC.  Since the mixing
  angles $\alpha$ and $\beta$ are close to the decoupling limit, the $H$
  is mainly produced via gluon fusion.  Hence a low mass $H$, where
  $H$ decays predominantly into $b\bar{b}$, will be difficult to
  observe at the LHC.  However, for a heavy $H$ the decay $H\to AZ$
  becomes available and could lead to  spectacular decay signatures of one
  photon and 4 leptons or two photon and leptons if the $A$ decays
  into $\gamma Z$ or $\gamma\gamma$ and the  $Z$'s decay
  leptonically.

  A low mass pseudo scalar $A$ has an enhanced production
  cross-section compared to the SM Higgs boson counterpart.  Moreover,
  its branching ratios into $\gamma Z$ and $\gamma\gamma$ are also
  enhanced due to the absence of massive gauge boson decay channels.
  This leads to a large signal strength for the $\gamma Z$ and
  $\gamma\gamma$ channels.   Therefore, current LHC data
    can exclude (or establish) this scenario.

Lastly we emphasize that our framework assumes 2HDM-II to be an
effective low energy theory of a more complete theory.  Our analysis
suggests the 2HDM-II description of the electroweak symmetry breaking
is only valid up to a scale around 1 TeV.  Hence, if nature chooses to
break electroweak symmetry by the 2HDM-II, there must be new particles
waiting to be discovered with masses  around 1 TeV.


\section*{Acknowledgements}
In the final stages of this work Ref.~\cite{Celis:2013rcs,Chiang:2013ixa,Barroso:2013awa} appeared. Our
work is complementary to those work.  We thank Jure Zupan for a critical reading of the manuscript.
PU thanks the hospitality of theCERN Theory Division where part of this work is being completed.  
The work of BG is supported by the U.S. Department of Energy under
contract No. DOE-FG03-97ER40546.  PU is supported by DOE grant
FG02-84-ER40153.


\appendix
\section{Loop-induced Decays of Neutral Scalar Bosons}
\label{app:decays}
The expressions for loop-induced decays of the Standard Model Higgs
boson, as well as for  the MSSM neutral bosons decays, are given in
Ref.~\cite{Djouadi:2005gi,Djouadi:2005gj}.  Here we give the rescaling
factors for the partial decay width suitable for the 2HDM.

The rescaling factors for $\phi\to gg$, $\gamma\gamma$ where $\phi=h$, $H$ are
\begin{equation}
\begin{aligned}
	\frac{\Gamma_{gg}}{\Gamma_{gg}^{SM}} &= \left|\frac{\sum_f c_t^\phi A_{1/2}(\tau_f)}{\sum_f A_{1/2}(\tau_f)}\right|^2,\\
	\frac{\Gamma_{\gamma\gamma}}{\Gamma_{\gamma\gamma}^{SM}} &= \left|\frac{\sum_f N_C Q_f^2c_f^\phi A_{1/2}(\tau_f)-c_V^\phi A_1(\tau_W)+\lambda_{\phi H^+H^-}A_0(\tau_{H^\pm})}{\sum_f N_C Q_f^2A_{1/2}(\tau_f) - A_1(\tau_W)}\right|^2,
\end{aligned}
\end{equation}
where $N_C$ and $Q_f$ are the number of colors and the electric charge
of fermion $f$.  $\lambda_{\phi H^+H^-}$ is the trilinear coupling of
$\phi$ to the charged Higgs boson in units of $2M_\phi^2/v$. The  functions
$A_i$ give the 1-loop contribution of a spin-$i$ particle.  They
are given by
\begin{equation}
\begin{aligned}
	A_0(\tau) &= -\frac{1}{\tau^2}\left(\tau-f(\tau)\right),\\
	A_{1/2}(\tau) &= \frac{2}{\tau^2}\left(\tau+(\tau-1)f(\tau)\right),\\
	A_1(\tau) &= \frac{1}{\tau^2}\left(2\tau^2+3\tau+3(2\tau-1)f(\tau)\right),
\end{aligned}
\end{equation}
and the function $f(\tau)$ is defined as
\begin{equation}
	f(\tau) = 
	\begin{cases}
		\arcsin^2\sqrt{\tau} & \tau\le1,\\
		-\frac{1}{4}\left(\log\frac{1+\sqrt{1-1/\tau}}{1-\sqrt{1-1/\tau}}-i\pi\right)^2 &\tau>1,
	\end{cases}
\end{equation}
with $\tau_i = M_\phi^2/4M_i^2$ for $i = f$, $W$ and $H^\pm$.

The rescaling factor for $\phi\to\gamma Z$ where $\phi=h$, $H$ is
\begin{equation}
\begin{scriptsize}
	\frac{\Gamma_{\gamma Z}}{\Gamma_{\gamma Z}^{SM}} = \left|\frac{\sum_f N_C\frac{Q_f(2I_f^3-4Q_fs_W^2)}{c_W}c_t^\phi A^\phi_{1/2}(\tau_f,\lambda_f)+c_V^\phi A^\phi_1(\tau_W,\lambda_W)+\frac{c_W^2-s_W^2}{c_W}\lambda_{\phi H^+H^-}A^\phi_0(\tau_{H^\pm},\lambda_{H^\pm})}{\sum_f N_C\frac{Q_f(2I_f^3-4Q_fs_W^2)}{c_W}A^\phi_{1/2}(\tau_f,\lambda_f)+ A^\phi_1(\tau_W,\lambda_W)}\right|^2,
\end{scriptsize}
\end{equation}
where $s_W(c_W)$ is the sine (cosine) of the weak mixing angle, $I_f^3$ is the left-hand weak isospin of the fermion $f$, $\tau_i=4M_i^2/M_\phi^2$ and $\lambda_i=4M_i^2/M_Z^2$.  
The function $A_i$'s are defined as
\begin{equation}
\begin{aligned}
	A^\phi_0(\tau,\lambda) &=  I_1(\tau,\lambda),\\
	A^\phi_{1/2}(\tau,\lambda) &= I_1(\tau,\lambda)-I_2(\tau,\lambda),\\
	A^\phi_1(\tau,\lambda) &= c_W\left\{4\left(3-\frac{s_W^2}{c_W^2}\right)I_2(\tau,\lambda)+\left[\left(1+\frac{2}{\tau}\right)\frac{s_W^2}{c_W^2}-\left(5+\frac{2}{\tau}\right)\right]I_1(\tau,\lambda)\right\},
\end{aligned}
\end{equation}
where the function $I_1$ and $I_2$ are defined as
\begin{equation}
\begin{aligned}
	I_1(\tau,\lambda) &= \frac{\tau\lambda}{2(\tau-\lambda)}+\frac{\tau^2\lambda^2}{2(\tau-\lambda)^2}\left(f(\tau^{-1})-f(\lambda^{-1})\right)+\frac{\tau^2\lambda}{(\tau-\lambda)^2}\left(g(\tau^{-1})-g(\lambda^{-1})\right),\\
	I_2(\tau,\lambda) &= -\frac{\tau\lambda}{2(\tau-\lambda)^2}\left(f(\tau^{-1})-f(\lambda^{-1})\right).
\end{aligned}
\end{equation}
The function $f$ is the same as in the $\phi\to\gamma\gamma$ case.  
The function $g$ is defined as
\begin{equation}
	g(\tau) = 
	\begin{cases}
		\sqrt{\tau^{-1}-1}\arcsin\sqrt{\tau} & \tau\ge1,\\
		\frac{\sqrt{1-1/\tau}}{2}\left(\log\frac{1+\sqrt{1-1/\tau}}{1-\sqrt{1-1/\tau}}-i\pi\right) &\tau<1.
	\end{cases}
\end{equation}

For the CP-odd boson, its loop induced partial decay width into $gg$, $\gamma\gamma$ and $\gamma Z$ are given by~\cite{Djouadi:2005gj}
\begin{equation}
\begin{aligned}
	\Gamma(A\to gg) &\simeq\frac{G_F\alpha_{s}^2 M_A^2}{36\sqrt{2}\pi^3}\left|\frac{3}{4}\cot\beta\frac{8M_t^2}{M_A^2}f(M_A^2/4M_t^2)\right|^2,\\
	\Gamma(A\to \gamma\gamma) &\simeq\frac{G_F\alpha_{em}^2 M_A^2}{128\sqrt{2}\pi^3}\left|\frac{4}{3}\cot\beta\frac{8M_t^2}{M_A^2}f(M_A^2/4M_t^2)\right|^2,\\
	\Gamma(A\to \gamma Z) &\simeq\frac{G^2_FM_W^2\alpha_{em} M_A^3}{16\pi^4}\left(1-\frac{M_Z^2}{M^2_A}\right)^3\left|2\cot\beta\frac{1-\frac{8}{3}s_W^2}{c_W}\frac{8M_t^2}{M_A^2}f(M_A^2/4M_t^2)\right|^2.
\end{aligned}
\end{equation}

\section{Beta-functions}
\label{app:beta}
Here we list the one-loop beta-function for the gauge, Yukawa and scalar couplings.  For the Yukawa couplings, we consider only the third generation contributions.  We use the shorthand notation $\alpha_i = g_i^2/(4\pi)$, $\alpha_y = y^2/(4\pi)$ and $\alpha_\lambda = \lambda/(4\pi)$.  The gauge coupling beta-functions are
\begin{equation}
	\beta_{\alpha_1} = \frac{7\alpha_1^2}{2\pi},\qquad
	\beta_{\alpha_2} = -\frac{3\alpha_2^2}{2\pi},\qquad
	\beta_{\alpha_3} = -\frac{7\alpha_3^2}{2\pi}.
\end{equation}
The Yukawa beta-functions are
\begin{equation}
\begin{split}
	\beta_{\alpha_{yt}} &= \frac{\alpha_{y_t}}{2\pi}\left(\frac{9}{2}\alpha_{y_t} - \frac{3}{2}\alpha_{y_b} - \frac{17}{2}\alpha_1 - \frac{9}{4}\alpha_2 - 8\alpha_3\right)\\
	\beta_{\alpha_{yb}} &= \frac{\alpha_{y_b}}{2\pi}\left(\frac{9}{2}\alpha_{y_b} - \frac{3}{2}\alpha_{y_t} - \frac{17}{2}\alpha_1 - \frac{9}{4}\alpha_2 - 8\alpha_3\right).
\end{split}
\end{equation}
Finally the scalar couplings beta-functions are
\begin{equation}
\begin{split}
	\beta_{\alpha_{\lambda1}} &= \frac{1}{4\pi}\left(12\alpha_{\lambda_1}^2 + 4\alpha_{\lambda_3}^2 + 4\alpha_{\lambda_3}\alpha_{\lambda_4} + 2\alpha_{\lambda_4}^2 + 2\alpha_{\lambda_5}^2\phantom{\frac12}\right.\\
	&\hspace{1.5cm}\left.+\frac{3}{4}\left(\alpha_1^2+3\alpha_2^2+2\alpha_1\alpha_2\right)-3\alpha_{\lambda_1}(\alpha_1+3\alpha_2-4\alpha_{y_b})-12\alpha_{y_b}^2\right)\\
	\beta_{\alpha_{\lambda2}} &= \frac{1}{4\pi}\left(12\alpha_{\lambda_2}^2 + 4\alpha_{\lambda_3}^2 + 4\alpha_{\lambda_3}\alpha_{\lambda_4} + 2\alpha_{\lambda_4}^2 + 2\alpha_{\lambda_5}^2\phantom{\frac12}\right.\\
	&\hspace{1.5cm}\left.+\frac{3}{4}\left(\alpha_1^2+3\alpha_2^2+2\alpha_1\alpha_2\right)-3\alpha_{\lambda_2}(\alpha_1+3\alpha_2-4\alpha_{y_b})-12\alpha_{y_b}^2\right)\\
	\beta_{\alpha_{\lambda3}} &= \frac{1}{4\pi}\left(2(\alpha_{\lambda_1}+\alpha_{\lambda_2})(3\alpha_{\lambda_3}+\alpha_{\lambda_4}) + 4\alpha_{\lambda_3}^2 +  2\alpha_{\lambda_4}^2 + 2\alpha_{\lambda_5}^2\phantom{\frac12}\right.\\
	&\hspace{1.5cm}\left.+\frac{3}{4}\left(\alpha_1^2+3\alpha_2^2+2\alpha_1\alpha_2\right)-3\alpha_{\lambda_3}(\alpha_1+3\alpha_2-2\alpha_{y_t}-2\alpha_{y_b})\right)\\
	\beta_{\alpha_{\lambda4}} &= \frac{1}{4\pi}\left(2(\alpha_{\lambda_1}+\alpha_{\lambda_2})\alpha_{\lambda_4} + 8\alpha_{\lambda_3}\alpha_{\lambda_4}+  4\alpha_{\lambda_4}^2 + 8\alpha_{\lambda_5}^2\phantom{\frac12}\right.\\
	&\hspace{1.5cm}\left.\phantom{\frac{3}{4}}+3\alpha_1\alpha_2 - 3\alpha_{\lambda_4}(\alpha_1+3\alpha_2-2\alpha_{y_t}-2\alpha_{y_b})\right)\\
	\beta_{\alpha_{\lambda5}} &= \frac{1}{4\pi}\left(2(\alpha_{\lambda_1}+\alpha_{\lambda_2})\alpha_{\lambda_5} + 8\alpha_{\lambda_3}\alpha_{\lambda_5}+  12\alpha_{\lambda_4}\alpha_{\lambda_5}\phantom{\frac12}\right.\\
	&\hspace{1.5cm}\left.\phantom{\frac{3}{4}} - 3\alpha_{\lambda_5}(\alpha_1+3\alpha_2-2\alpha_{y_t}-2\alpha_{y_b})\right)\\
\end{split}
\end{equation}
\bibliography{2hdm}
\bibliographystyle{JHEP}

\end{document}